\documentclass[english,aps,prb,twocolumn, preprintnumbers,amsmath,amssymb]{revtex4}

\usepackage[T1]{fontenc}
\usepackage[latin9]{inputenc}
\usepackage{color}
\usepackage{units}
\usepackage{amssymb}
\usepackage{graphicx}
\usepackage{esint}
\usepackage{bm}
\usepackage{natbib}
\usepackage{ulem}
\usepackage[colorlinks=true, citecolor=red]{hyperref}
\setcitestyle{journalcolor= blue}

\usepackage{graphicx}
\usepackage{dcolumn}
\usepackage{bm}

\usepackage{babel}

\begin{document}

\title{Probing defect states in few-layer MoS$_{2}$ by conductance fluctuation spectroscopy}

\author{Suman Sarkar}
\affiliation{Department of Physics, Indian Institute of Science, Bangalore 560012, India}
\author{ K. Lakshmi Ganapathi }
\affiliation{Department of Physics, Indian Institute of Technology, Madras, India}
\author{ Sangeneni Mohan }
\affiliation{Centre for NanoScience and Engineering, Indian Institute of Science, Bangalore-560012, 	India}
\author{Aveek Bid}
\email{aveek@iisc.ac.in}
\affiliation{Department of Physics, Indian Institute of Science, Bangalore 560012, India}

\begin{abstract}

Despite the concerted effort of several research groups, a detailed experimental account of defect dynamics in high-quality single- and few-layer transition metal dichalcogenides remain elusive.  In this paper we report an experimental study of  the temperature dependence of conductance and conductance-fluctuations on several few-layer MoS$_{2}$ exfoliated on hexagonal boron nitride and covered by a  {capping layer of} high-$\kappa$ dielectric HfO$_{2}$. The presence of the high-$\kappa$ dielectric made the device extremely stable against environmental degradation as well as resistant to changes in device characteristics upon repeated thermal cycling  enabling us to obtain reproducible data on the same device over a time-scale of more than one year. Our device architecture helped bring down the conductance fluctuations of the MoS$_2$ channel by orders of magnitude  compared to previous reports. The extremely low noise levels in our devices made in possible to detect the generation-recombination noise arising from charge fluctuation between the sulphur-vacancy levels in the band gap and energy-levels  at the conductance band-edge. Our work establishes conduction fluctuation spectroscopy as a viable route to quantitatively probe in-gap defect levels in low-dimensional  semiconductors. 

\end{abstract}

\maketitle
Following the discovery of graphene~\cite{novoselov2004electric}, the exploration of the basic physics and technological implications of two-dimensional (2D) materials has gained tremendous importance. Though graphene is a system rich in novel physics, the lack of band-gap limits its applications in transistor technology. Transition metal dichalcogenides (TMD) like MoS$_{2}$ and MoSe$_2$, on the other hand, have band-gaps of the order of  eV in the few-layer limit~\cite{mak2010atomically} making them ideal for opto-electronic applications ~\cite{wang2012electronics,radenovic2011single,eda2013two}. On the flip-side, the reported mobilities of these TMD based field effect transistor (FET) devices are very low ~\cite{doi:10.1021/ar4002312, radenovic2011single} and the quoted values vary widely between samples. It is now understood that  defect-levels (primarily arising from chalcogenide vacancies)  adversely affect the mobilty and optical properties of  these TMD-based devices~\cite{hong2015exploring,ong2013mobility,qiu2013hopping,zhu2014electronic}. 

Despite extensive research, there is no clear understanding of the underlying defect dynamics in this system. Traditional transport measurements like current-voltage characteristics and the temperature dependence of the resistance, while providing indications of the existence of defect states,  cannot directly probe their energetics ~\cite{qiu2013hopping,mcdonnell2014defect}. Photoluminescence measurements report the appearance, at low temperatures of an additional peak in the spectrum which is tentatively attributed to transitions from a `defect'-level ~\cite{tongay2013defects,saigal2016evidence}, but no direct evidence of this level has been found from optical studies. Transmission electron microscopy (TEM) ~\cite{hong2015exploring,lin2014properties,zou2012predicting,van2013grains,najmaei2013vapour} and scanning tunnelling microscopy (STM) ~\cite{hosoki1992surface,permana1992observation,ha1994scanning,murata2001scanning,kodama2010electronic,addou2015surface} have shown that the primary point-defects are  $\mathrm{S}$-vacancies although other types of defects like interstitials, dislocations, dopants and grain boundaries were also seen. These two techniques come with their own sets of limitations. While TEM imaging is believed to induce additional defects in MoS$_2$ ~\cite{komsa2012two,komsa2013point},  atomic-resolution  imaging of few-layer TMD using STM  has proved challenging ~\cite{mcdonnell2014defect,lu2014bandgap, huang2015bandgap, vancso2016intrinsic}. Thus, although theoretical studies predict the presence of prominent defect-levels in these materials~\cite{KC_2014, PhysRevMaterials.2.084002, krivosheeva2015theoretical,lin2016defect,noh2014stability}, probing them  experimentally   has proved to be challenging. 

In this paper, we present conductance fluctuation spectroscopy~\cite{RevModPhys.53.497} as a viable technique to identify these defect states and their characteristic energy levels.  Conductance fluctuations (noise) in TMD-based devices has been studied by several groups~\cite{song2017probing,kwon2014analysis,sangwan2013low,ghatak2014microscopic,renteria2014low}. In different studies, the  observed conductance fluctuations have been variously attributed to  charge-carrier number density fluctuations  due to trapping at the interface~\cite{kwon2014analysis}, to mobility fluctuations~\cite{sangwan2013low,ghatak2014microscopic} or to contact noise~\cite{renteria2014low}. In general, in the high doping regime, carrier-number density fluctuation model could explain the measured noise behavior while in the low doping regime  mobility fluctuation models seemed to better fit  the experimental observations~\cite{na2014low}. Thus there is a lack of consensus in the community as to the origin of the observed large conductance fluctuations in this system.   The problem is aggravated by the fact that ultra-thin layers of TMD degrade extremely fast when exposed to the ambient~\cite{renteria2014low,qiu2012electrical,doi:10.1021/nn301572c,kooyman2008detrimental}. This makes repeated, reliable measurements on the same device challenging while at the same time severely limiting the scope of practical applications. 

We have performed  detailed measurements of temperature $T$ dependence of conductance and conductance-fluctuations on several few-layer MoS$_{2}$ exfoliated on hexagonal boron nitride (hBN) and{covered by} a film of high-$\kappa$ dielectric HfO$_{2}$. We find that over a large range of $T$, the noise in the system is dominated by  generation-recombination processes caused by random  charge fluctuations via transitions between the $\mathrm{S}$-vacancy  impurity band and the conduction band of MoS$_2$.  The{presence of the HfO$_2$ capping-layer} makes it extremely stable against degradation upon exposure to the atmosphere and to repeated thermal-cycling. The presence of the crystalline hBN beneath  screens the device from charge-fluctuations in the SiO$_2$ substrate resulting in the noise levels in our device being orders of magnitude smaller than previous reports of on-substrate devices. This enabled us to detect charge fluctuations between the  $\mathrm{S}$-vacancy levels and  the conduction-band edge. 

\begin{figure}[t]
	\begin{center}
		\includegraphics[width=0.5\textwidth]{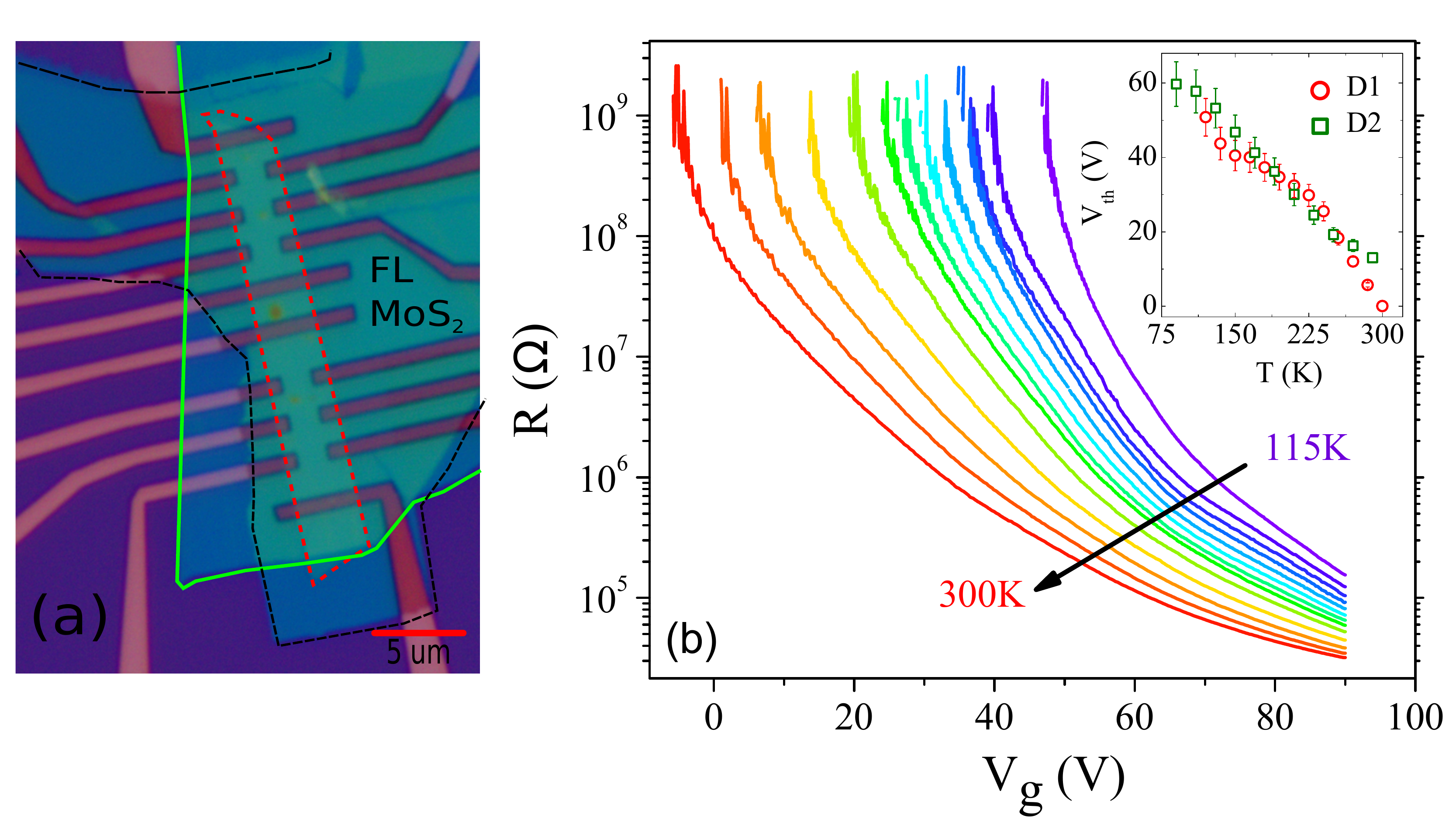}
		\small{\caption{(a) Optical image of device D1. The bottom hBN is defined with a solid green line and the few-layer (FL) MoS$_{2}$ is outlined by a red  dashed line. The top HfO$_{2}$ is outlined by a dashed black line. (b) Gate-voltage $V_g$ dependence of the resistance $R$ of device D1 at a few representative temperatures ranging from 115 to 300~K in steps of 15~K. The inset plots the on-set voltage $V_{th}$ versus $T$ for   devices D1 (red open circles) and D2 (green open squares). \label{fig:1}}}
	\end{center}
\end{figure}

Samples were prepared in FET configuration by conventional Polydimethylsiloxane (PDMS) assisted dry-transfer method~\cite{dean2010boron}. We studied two classes of devices. In the first class, a few-layer hBN flake ($\approx~20$~nm thick) was transferred on Si$^{++}$/SiO$_{2}$ substrate followed by the transfer of a few-layer MoS$_{2}$  on top. The transfers were made using a custom-built set-up based on a motorized XYZ-stage (Thorlabs model B51x) using a long working distance 50X-objective under an optical microscope. Electrical contacts were defined by standard electron-beam lithography followed by thermal deposition of 5~nm Cr and 25~nm Au. This was followed by an electron-beam assisted evaporation of 30~nm of HfO$_{2}$ covering the entire surface of the device. The  HfO$_{2}$  thin film was  deposited directly  on MoS$_{2}$ without any buffer layer or surface treatment - the details of the HfO$_2$ film growth  are discussed elsewhere~\cite{lakshmi2013optimization,ganapathi2014influence}. Several such devices were tested. In this article we concentrate on the results obtained on one such device, labeled D1. For comparison, we also studied a second class of devices - these were  few-layer MoS$_2$ devices fabricated directly on the Si$^{++}$/SiO$_2$ substrate without the top encapsulation layer (labeled D2). The thickness of the SiO$_2$ in all cases was 295~nm.    {In all cases, the gate bias voltage, $V_g$ is applied from underneath Si$^{++}$/SiO$_2$ substrate.} 

Electrical transport measurements were performed in a two-probe configuration using low-frequency lock-in technique. The bias voltage across the device was set to $V_{ds}=$~5~mV. The current $I_{ds}$ flowing through the device was amplified by a low-noise current-amplifier (Ithaco~1211) and measured by a digital dual-channel Lock-in-amplifier (LIA).  The gate voltage, $V_g$ was controlled by a   Keithley-2400 source-meter. 

An optical image of the device D1 is shown in Fig.~\ref{fig:1}(a), the few-layer MoS$_2$ (encapsulated between hBN and HfO$_2$) is outlined by a black dashed line.  Fig.~\ref{fig:1}(b) shows a plot of the sheet resistance $R\equiv V_{ds}/I_{ds}$ of the device D1 versus $V_g$ measured over the temperature range 115--300~K. The gate response of the device establishes it to be an n-type semiconductor which is typically what is observed in naturally occurring  MoS$_{2}$. The large on-off ratio ($\sim10^{5}$), low on-state resistance  ($\sim30$~K$\Omega$)  and very low off-state current ($\sim$10~pA) attest to the high-quality of the device. From the inset of Fig.~\ref{fig:1}(b) it can be seen that the threshold voltage $V_{th}$  decreases sharply with increasing temperature going to negative $V_g$  near room-temperature. On the other hand, $V_{th}$ for D2 at room temperature was $\sim15$~V.

 \begin{figure}[t]
	\begin{center}
		\includegraphics[width=0.5\textwidth]{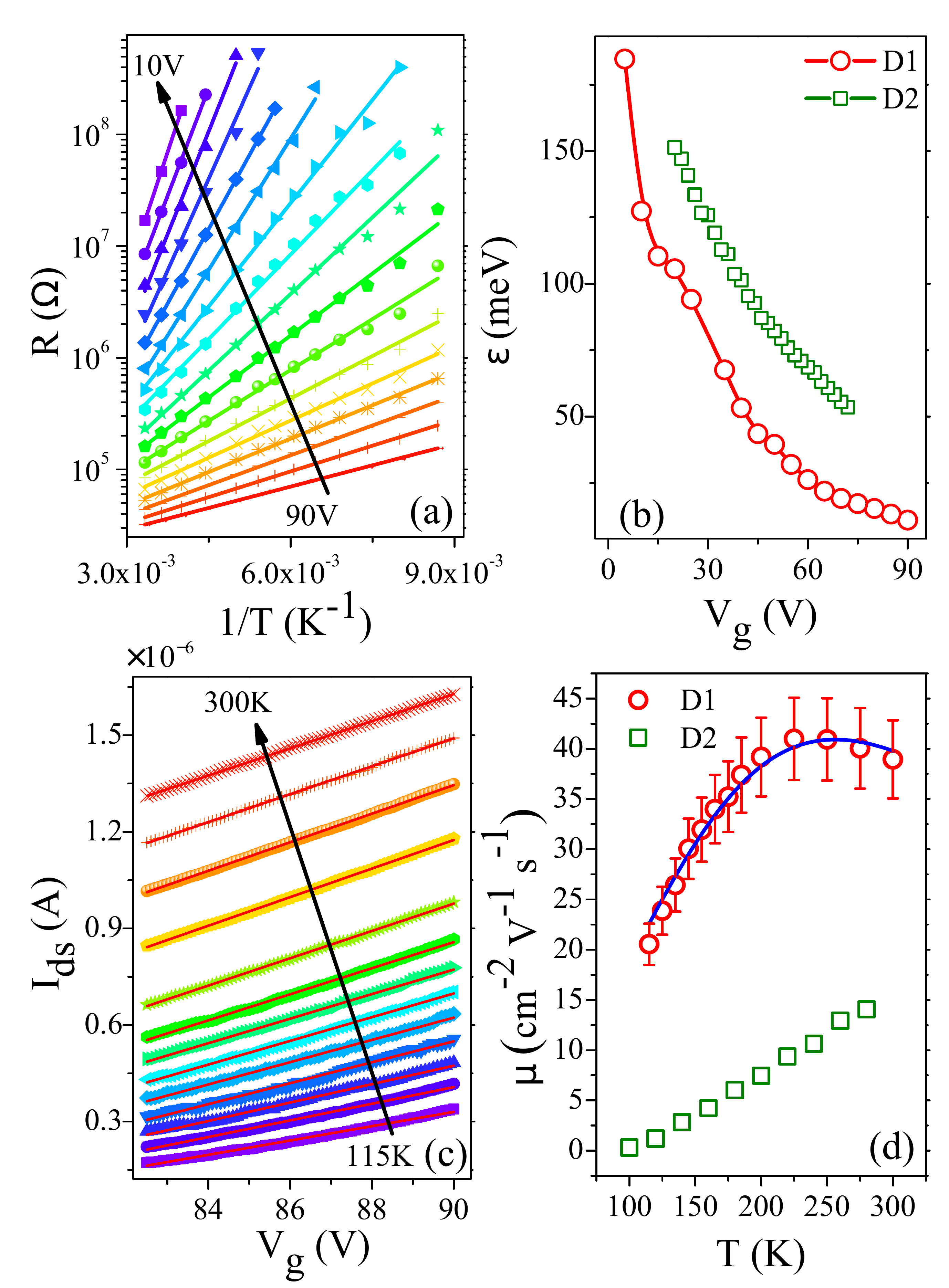}
		\small{\caption{(a) Scatter plot of resistance $R$ of device D1 plotted on a semi-logarithmic scale versus $1/T$ at several representative values of $V_g$ from 10~V to 90~V in steps of  5~V. The solid lines are the linear fits of $1/T$ vs ln($R$). (b) Plots of activation energy $\varepsilon$ versus $V_g$ extracted from the ln($R$) versus $1/T$ plots for the two devices, D1 (red open circles) and D2 (green open squares). The lines are guides to the eye. (c) Plots of   $I_{ds}$ versus $V_g$  at different $T$ ranging from 115~K to 300~K for device D1. (d) Plots of mobility $\mu$ versus $T$ for the two devices, D1 (red open circles) and D2 (green open squares). The blue line is fit to the data for D1 using Eqn.~\ref{Eqn:mu}.  \label{fig:2}}}
	\end{center}
\end{figure} 

In Fig.~\ref{fig:2}(a) we plot  the sheet resistance $R$ of the device D1 in a semi-logarithmic scale versus inverse temperature for few representative values of $V_g$. The linearity of the plots indicates that, at least in the high $T$ limit, electrical transport is dominated by thermal activation of the charge carriers. More specifically,  as we go higher in $V_g$, the range of $T$ where this linearity holds extends down to lower temperatures. The activation energy, $\varepsilon$ extracted from the slope of the $ln(R)$ versus $1/T$ plots is plotted in Fig.~\ref{fig:2}(b). One can see that $\varepsilon$ increases as one decreases the gate bias and it varies from  20~meV at high $V_g$ to an order of magnitude higher $\sim$200~meV,  close to the off-state of the device. The activation energy for device D2, extracted in a similar fashion is, as expected,  higher than that of D1 at all values of $V_g$. 

\begin{figure}[t]
	\begin{center}
		\includegraphics[width=0.5\textwidth]{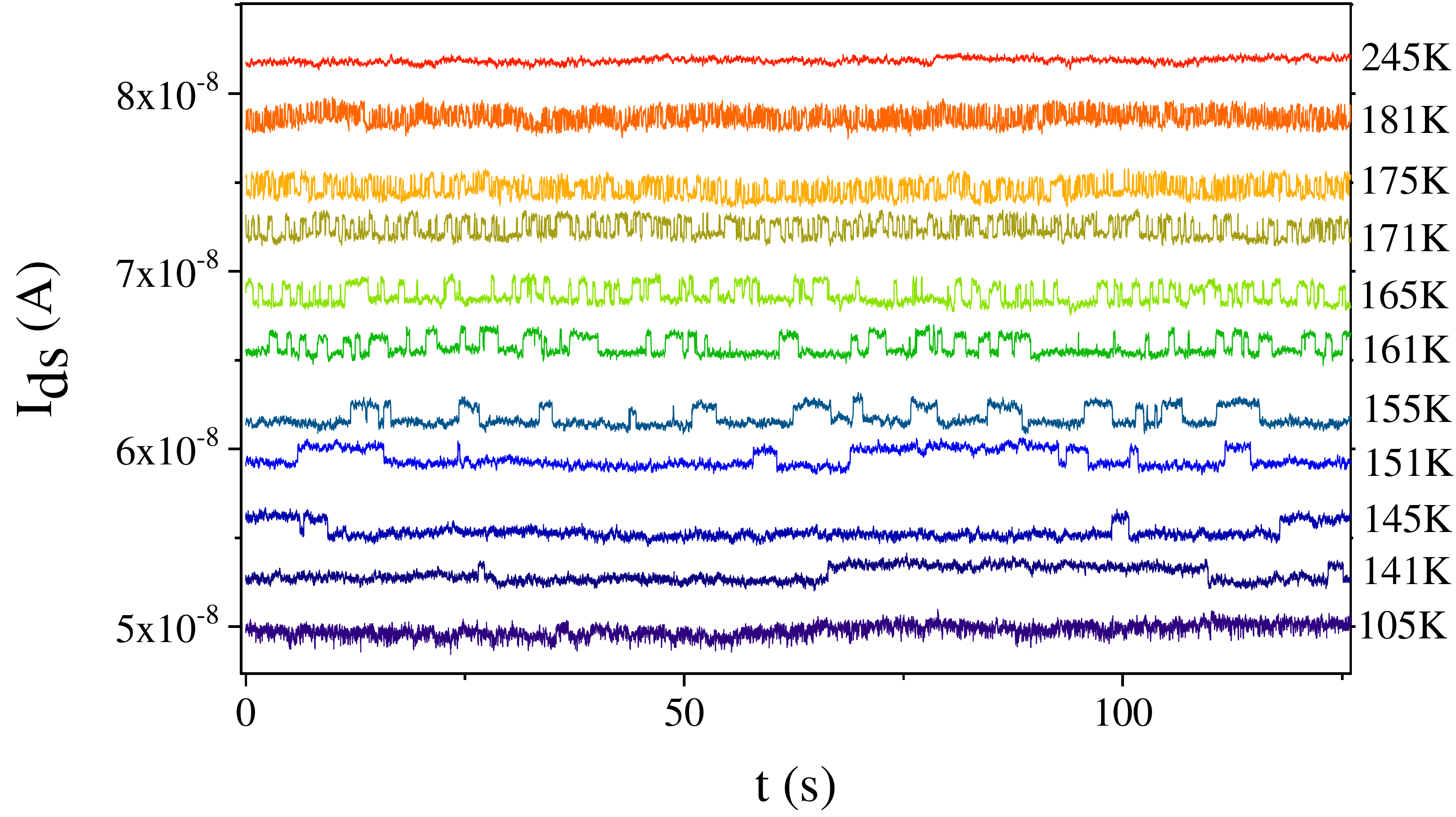}
		\small{{\caption{Plots of source-drain current $I_{ds}$ versus time at a few representative temperatures from 105~K to 245~K for device D1.  The data were taken for $V_g=90$~V. \label{fig:3}}}}
	\end{center}
\end{figure}

The field-effect mobility $\mu$ of the devices can be obtained from the relation $\mu=\frac{dI_{ds}}{dV_{g}}\frac{L}{wCV_{ds}}$. Here $L$ is the length of the channel, $w$ its width and $C$ is the gate capacitance per unit area. In Fig.~\ref{fig:2}(c) we show plots of $I_{ds}$  versus $V_g$, the slope of this curve gives the mobility of the device.  A plot of the $T$ dependence of the mobility is shown in Fig.~\ref{fig:2}(d). We find that $\mu$ for device D1 is $\sim20$~cm$^{-2}$V$^{-1}$s$^{-1}$ at 100~K. With increasing $T$, $\mu$ increases monotonically till about 225~K beyond which it begins to fall with increasing $T$. To understand the measured  $T$ dependence of $\mu$ we note that in  2D semiconductors, the mobility of the charge carriers is affected by Coulomb scattering, acoustic and optical phonon scattering, scattering by the interface phonon and roughness due to the surface~\cite{kim2012high}. At high--$T$, scattering due to phonons  is dominant which causes the mobility to have a $T^{-3/2}$ dependence~\cite{sangwan2018electronic,mao2017manipulation}. On the other hand, scattering from charge impurities located randomly in the sample is the dominant factor limiting $\mu$ at low temperatures causing the mobility to depend on temparature as $T^{3/2}$~\cite{ong2013mobility,liu2017unified}. Following Matthiessen's rule: 
\begin{eqnarray}
\frac{1}{\mu}=\frac{1}{M_pT^{-3/2}} + \frac{1}{M_iT^{3/2}}
\label{Eqn:mu}
\end{eqnarray}
where $M_p$ and $M_i$ represent the relative contributions of  the phonon-scattering and impurity-scattering mechanisms respectively.  These coefficients are not independent, but are related by $(M_p/M_i)^{1/3}=T_{max}$, where $T_{max}$ is the temperature at which  $\mu$ has a maxima. In Fig.~\ref{fig:2}(d) we show a fit of the $T$ dependence of the mobility of D1 to Eqn.~\ref{Eqn:mu}. The mobility of D2, on the other hand, monotonically increases with $T$ showing that over the  range of $T$ studied, impurity-scattering dominates the transport in on-SiO$_2$ substrate devices.

The presence of both bulk- and surface-transport channels complicates the charge transport in these systems. To understand the charge-carrier dynamics arising from the surface- and bulk-states in this system, we studied the low-frequency conductance fluctuations over the temperature range 70~K-300~K using a 2-probe ac digital-signal-processing technique~\cite{Aveekthesis}. As established in several previous reports, $1/f$ noise is an excellent parameter to probe inter-band scattering of charge-carriers in systems with multiple conduction channels ~\cite{daptary2018effect,price1981two,hooge19941}. We used an SR830 dual-channel digital LIA to voltage-bias the sample at a carrier frequency of $f_{0}\sim228~Hz$.  The current $I_{ds}$ through the device was amplified by the low-noise current preamplifier and detected by the LIA.  The data were acquired at every $T$ and $V_g$ for 32~minutes at  a sampling rate of 2048~points/s using a fast 16-bit data acquisition card. This time-series of current fluctuations $\delta I_{ds}(t)$ was digitally anti-alias filtered and decimated. The power spectral density (PSD) of current-fluctuations, $S_I(f)$  was calculated from this filtered time-series using the method of Welch-periodogram~\cite{Aveekthesis,scofield1987ac}.  The system was calibrated by measuring the thermal (Johnson-Nyquist) noise of standard resistors.  The time-series of $I_{ds}$ measured for device D1 at a few representative temperatures at $V_g=90$~V  are shown in Fig.~\ref{fig:3}. We find that over the $T$ range $\sim$140-190~K,  the measured $I_{ds}(t)$ (and consequently the conductance $g(t)=I_{ds}(t)/V_{ds}$) for D1 fluctuates between two well-defined levels~\cite{PhysRev.78.615,kundu2017quantum}. This `Random telegraphic noise' (RTN)~\cite{hung1990random} usually signifies that the system has access to two (or more) different states separated  by an energy barrier.  We come back later in this article  to a discussion of  the detailed statistics of the RTN and the physical origin of these states. 

\begin{figure}[t]
	\begin{center}
		\includegraphics[width=0.5\textwidth]{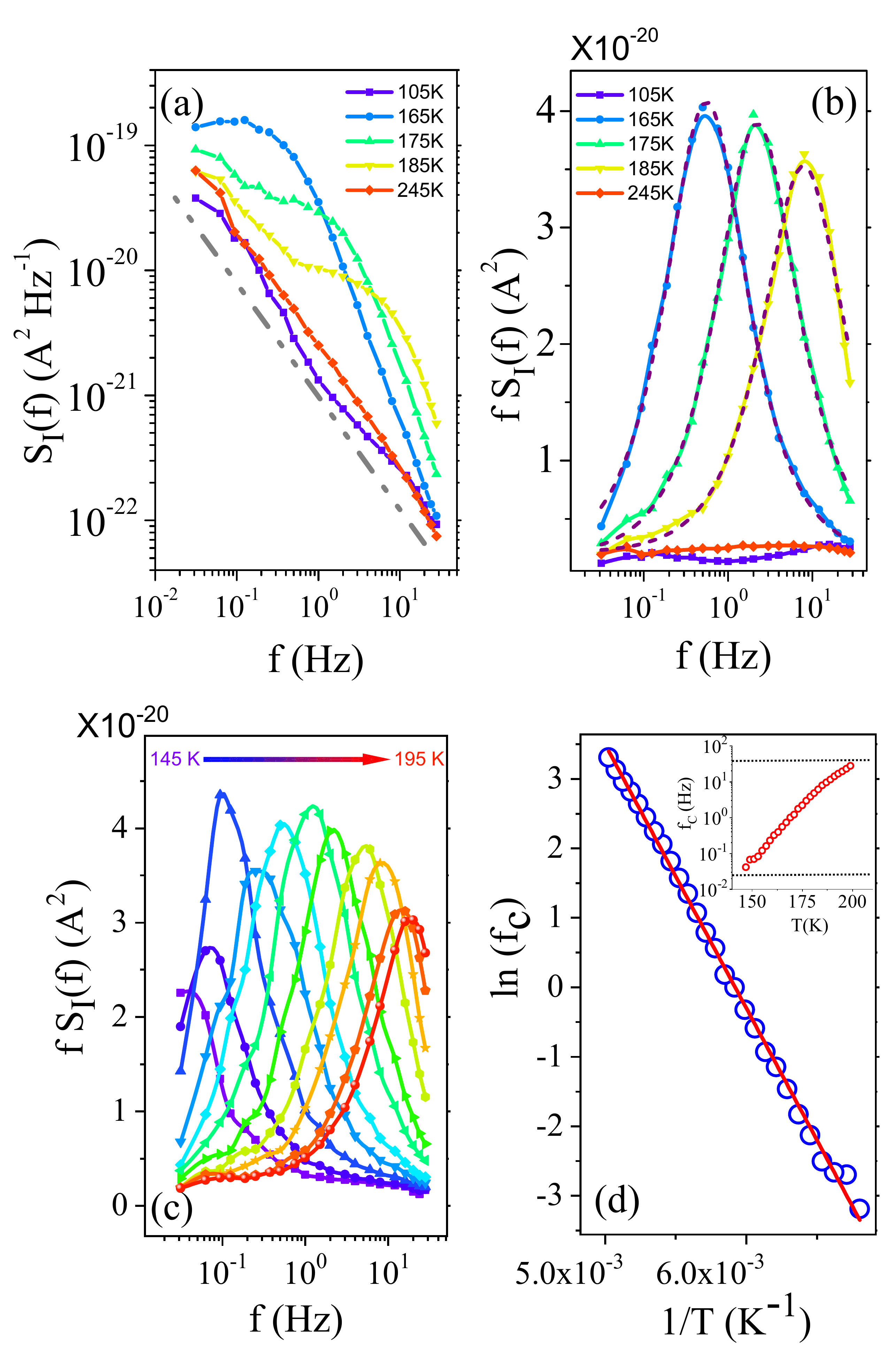}
		\small{{\caption{ (a) Plots of $S_I(f)$ versus $f$ at a few representative temperatures for device D1. The data were measured for $V_g=90$~V. (b) Plots of $fS_I(f)$ versus $f$ for the same values of $T$ as in  (a). The  dotted purple lines are  fits  using Eqn.~\ref{Eqn:2level} to the data at 165~K, 175~K  and 185~K. (c) Plot of $fS_I(f)$ as a function of $f$ over the temperature range 145~K (purple data points) --195~K (red data points) in steps of 5~K. The arrow indicates the evolution of $f_C$ to higher values with increasing $T$. The solid lines are guides to the eye. (d) Plot of the logarithm of $f_C$ versus $1/T$. The solid line is a linearized fit to the Arrhenius relation $f_C$= $f_0$$exp(-E_a/k_{B}T)$. The inset shows a plot of $f_C$ versus $T$.  The two dotted lines are the upper (28~Hz )and lower (31.25~mHz) limits of our measurement band-width.  \label{fig:4}}}}
	\end{center}
\end{figure}
\begin{figure}[t]
	\begin{center}
		\includegraphics[width=0.5\textwidth]{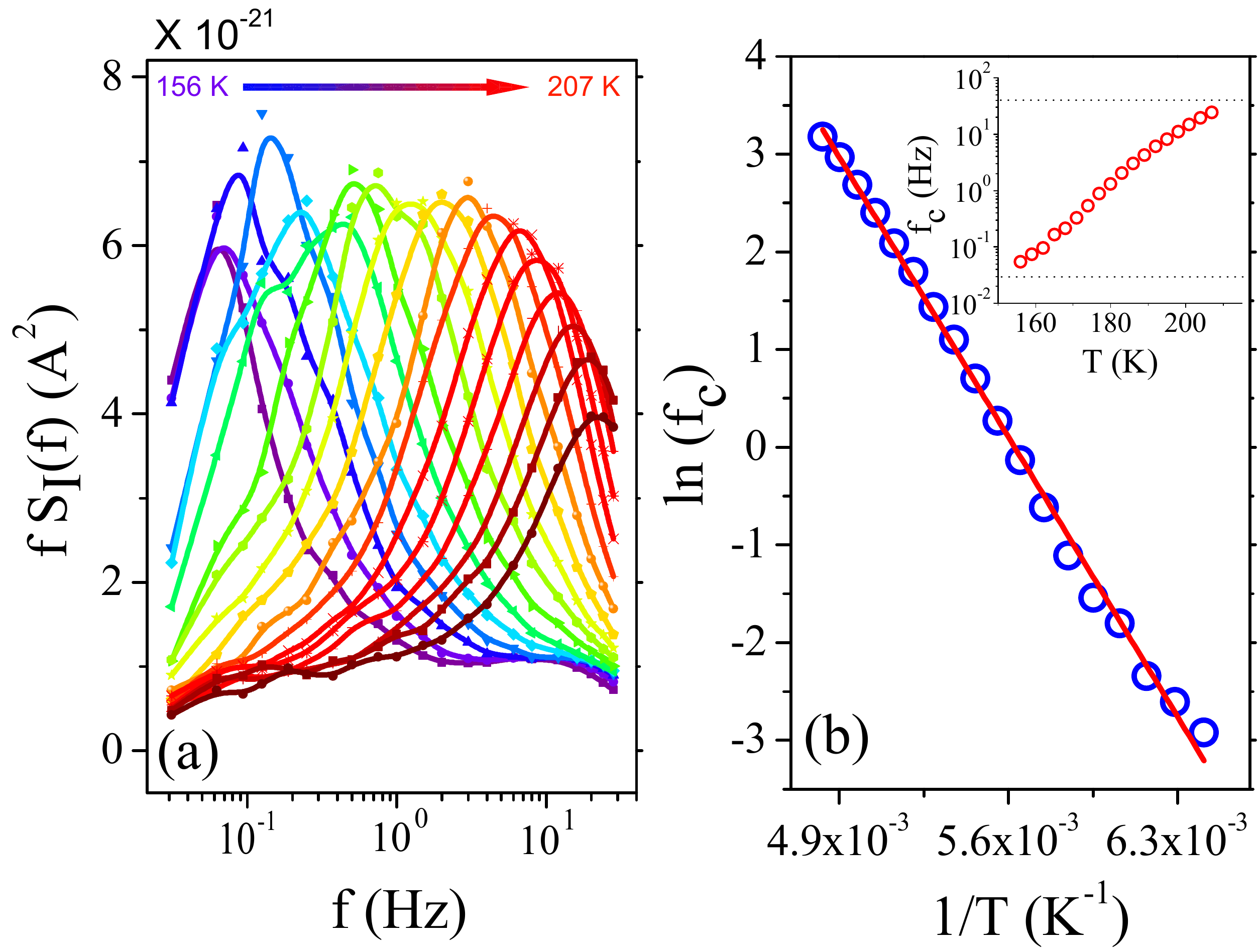}
		\small{{\caption{(a) Plot of $fS_I(f)$ as a function of $f$ over the temperature range 156~K (purple data points) -- 207~K (red data points) in steps of 4~K for the device D1b. (b) Plot of the logarithm of $f_C$ versus $1/T$. The solid line is a linearized fit to the Arrhenius relation $f_C$= $f_0$$exp(-E_a/k_{B}T)$. The inset shows a plot of $f_C$ versus $T$. The measurements were done at $V_g$ = 90~V. \label{fig:5}}}}
	\end{center}
\end{figure}

\begin{figure}[t]
	\begin{center}
		\includegraphics[width=0.5\textwidth]{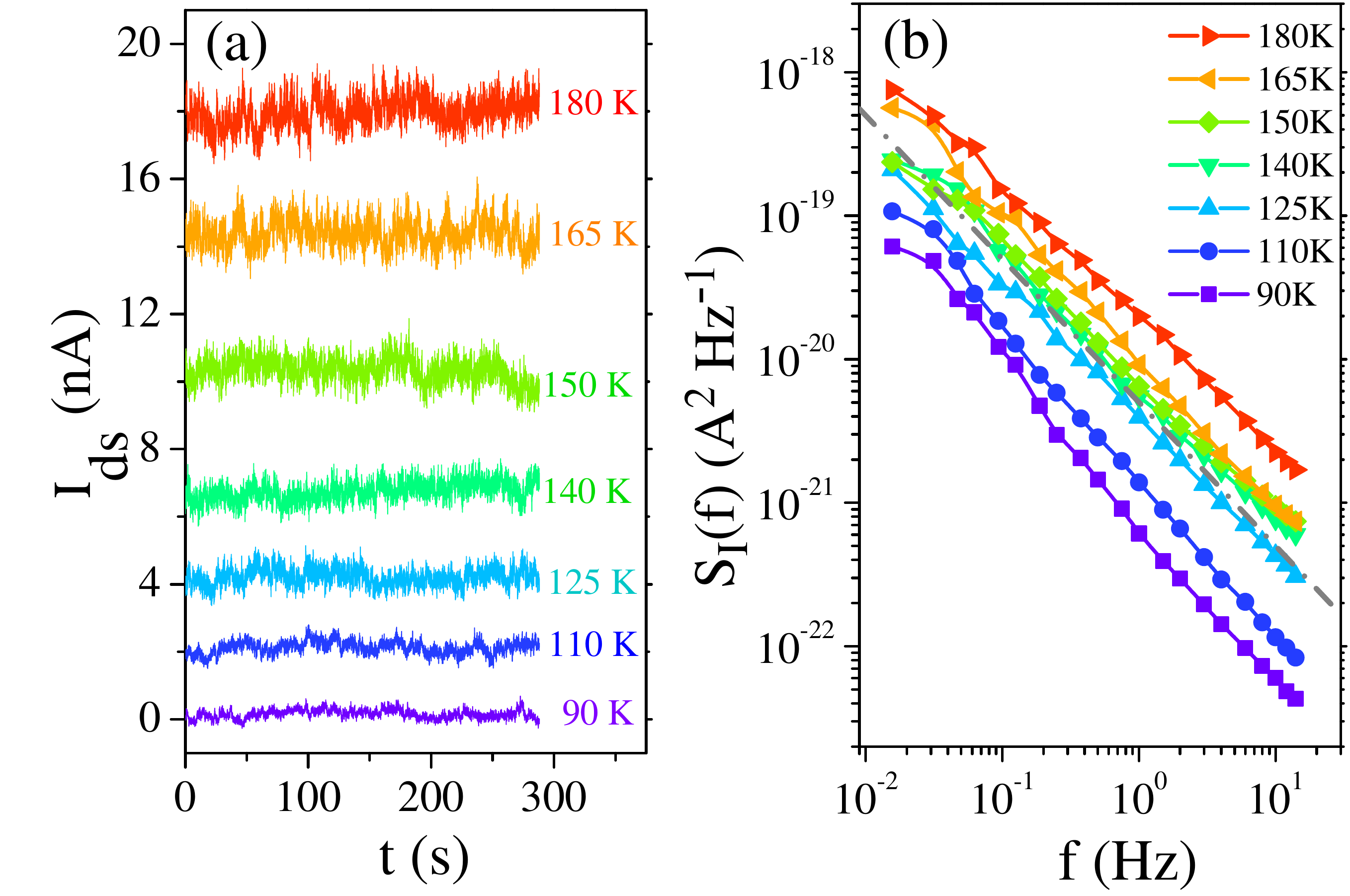}
		\small{{\caption {(a) Plot of $I_{ds}$ versus time at a few representative values of $T$ for the device D2. (b) PSD $S_I(f)$  corresponding to the time-series shown in (a).  The gray line shows a representative $1/f$ curve. The measurements were done at $V_g=72$~V.\label{fig:6}}}}
	\end{center}
\end{figure}	

In Fig.~\ref{fig:4}(a) we plot the PSD of current fluctuations at a few representative values of $T$ and $V_g=90$~V for the device D1. We find that over the $T$ range where RTN were present in the time-series $I_{ds}(t)$, the PSD deviates significantly from the  $1/f$ dependence (shown  in the plot by a gray line).  This can be appreciated better from Fig.~\ref{fig:4}(b) where we plot the quantity $fS_I(f)$ which should be independent of frequency for $1/f$ noise, as is indeed the case for the PSD measured at 105~K and 245~K. On the other hand, the PSD measured in the intermediate $T$ range ($140$~K$<T<190$~K) has a significant non-$1/f$ component. The PSD of an RTN is a Lorentzian with a characteristic frequency $f_C$, where $1/f_C=\tau_c$ is the typical time-scale of switching between the two distinct levels. This motivated us to fit the measured PSD of current fluctuations to an equation which contains both  $1/f$ and  Lorentzian components:
\begin{equation}
\frac{S_I(f)}{I^2} = \frac{A_1}{f} + \frac{A_2f_C}{f^2+f_C^2}
\label{Eqn:2level}
\end{equation}
$A_1$ and $A_2$ are fit parameters that denote the relative contributions of the random and RTN fluctuations respectively to the total PSD. The dotted purple lines are  fits to the data at 165~K, 175~K  and 185~K using Eqn.~\ref{Eqn:2level}.  In Fig.~\ref{fig:4}(c) we show plots of  $fS_I(f)$ versus $f$ over an extensive range of $T$. We find that as $T$ increases,  the peak position evolves from a few~mHz to few tens of Hz [see the inset of Fig.~\ref{fig:4}(d). Beyond this $T$ range, the value of $f_C$ goes beyond our measurement frequency bandwidth (31.25~mHz--28~Hz). The value of $f_C$ is thermally activated and follows the Arrhenius relation: $f_C$= $f_0$$exp(-E_a/k_{B}T)$.  Figure~\ref{fig:4}(d) shows a plot of ln($f_C$) versus $1/T$, the red-line is a fit to the activated behaviour. The value of activation energy $E_a$  extracted from the fit is 370~meV.  {These measurements were repeated on three such devices (MoS$_2$ encapsulated between hBN and HfO$_2$); we find that the activation energy-scale in all of them lie in the range $370±30$ meV. In Fig.~\ref{fig:5}, we show data for another device, D1b, for which we obtain $E_a$=353~meV.}  We come back to the physical implications of this energy-scale later in this article.  

The time-series $\delta I(t)$ for device D2, on the other hand,  did not have any RTN component (Fig.~\ref{fig:6}(a)) and the PSD had a $1/f^\alpha$ (with $0.9<\alpha<1.1$) dependence on $f$ over the entire $T$ and $V_g$ range studied (Fig.~\ref{fig:6}(b)).  {The $I_{ds}$(t) data were obtained for the device D2 at 72~V. We have attempted to compare the data in the two sets of devices at similar values of number-densities. Due to the presence of the hBN layer, the effective thickness of the dielectric layer in D1 was higher than that of D2 -- requiring a higher gate-voltage for D1 than that for D2 to achieve similar carrier number density.  On the other hand, D1 had a lower threshold voltage than D2. Taking both these factors into account, we have estimated the $V_g$ at which the induced number densities are similar for both D1 and D2.  Thus, for D1, the data are presented for $V_g$ =90~V while for the device D2, the data are presented for $V_g$ = 72~V.}
 
\begin{figure}[t]
	\begin{center}
		\includegraphics[width=0.5\textwidth]{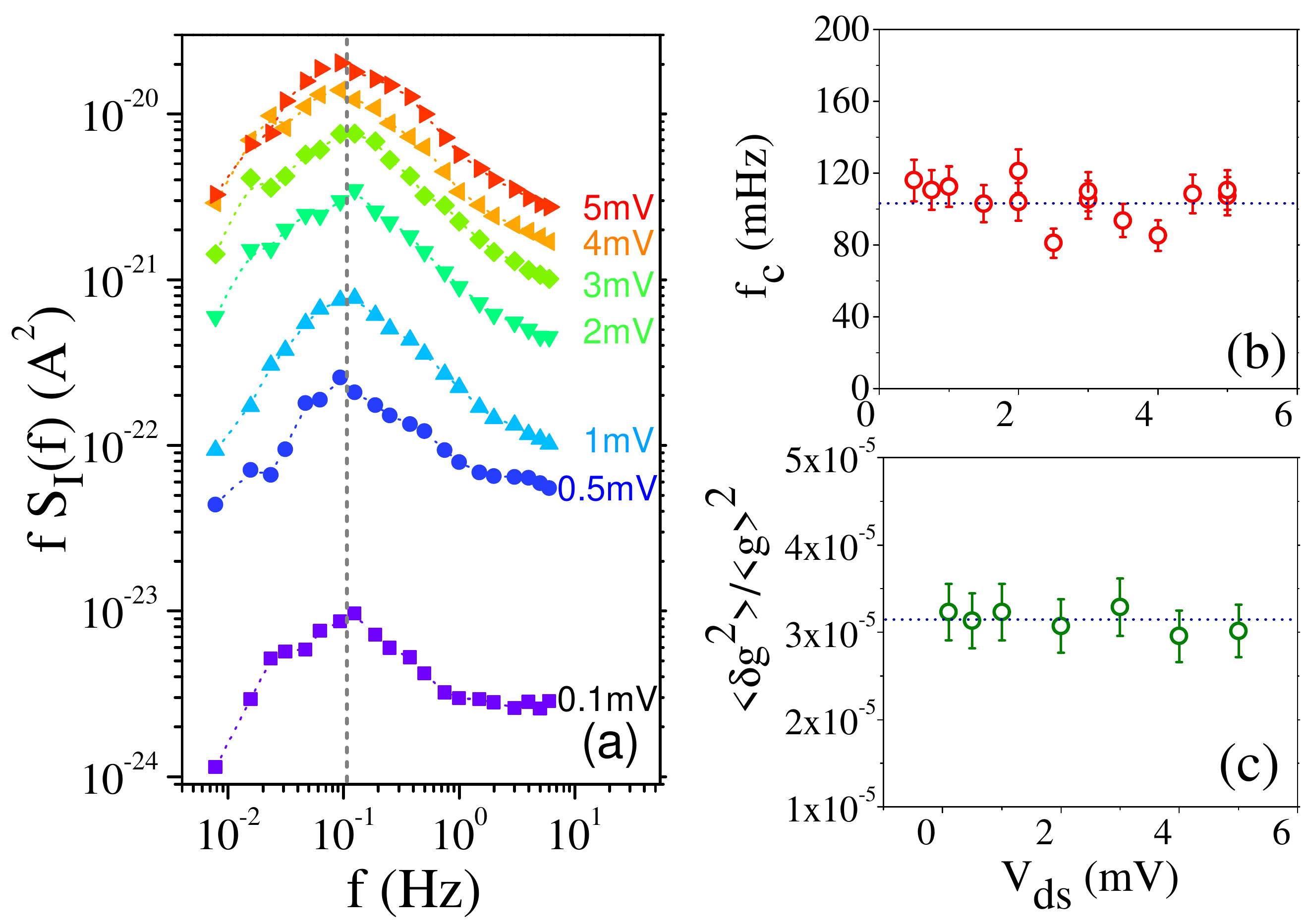}
		\small{{\caption{(a) Plot of $fS_I$(f) versus frequency $f$ at different source-drain bias $V_{ds}$ across the sample D1. The vertical dashed gray line is a visual guide indicating the positions of the Lorentzian peaks in the spectra. The value of $V_{ds}$ is marked next to each curve.  (b). Plot of  $f_{c}$ as a function of $V_{ds}$ extracted from the PSD in (a) using Eqn.~\ref{Eqn:2level}. The dotted line is a visual guide to the eye. (c) Plot of the relative variance of conductance fluctuations $\mathcal{G}_{var}$ as a function of $V_{ds}$, the dotted  line is a guide to the eye. The data were all taken at  $T$ = 175~K and $V_g$ = 90~V.  \label{fig:7}}}}
	\end{center}
\end{figure}
\begin{figure}[t]
	\begin{center}
		\includegraphics[width=0.5\textwidth]{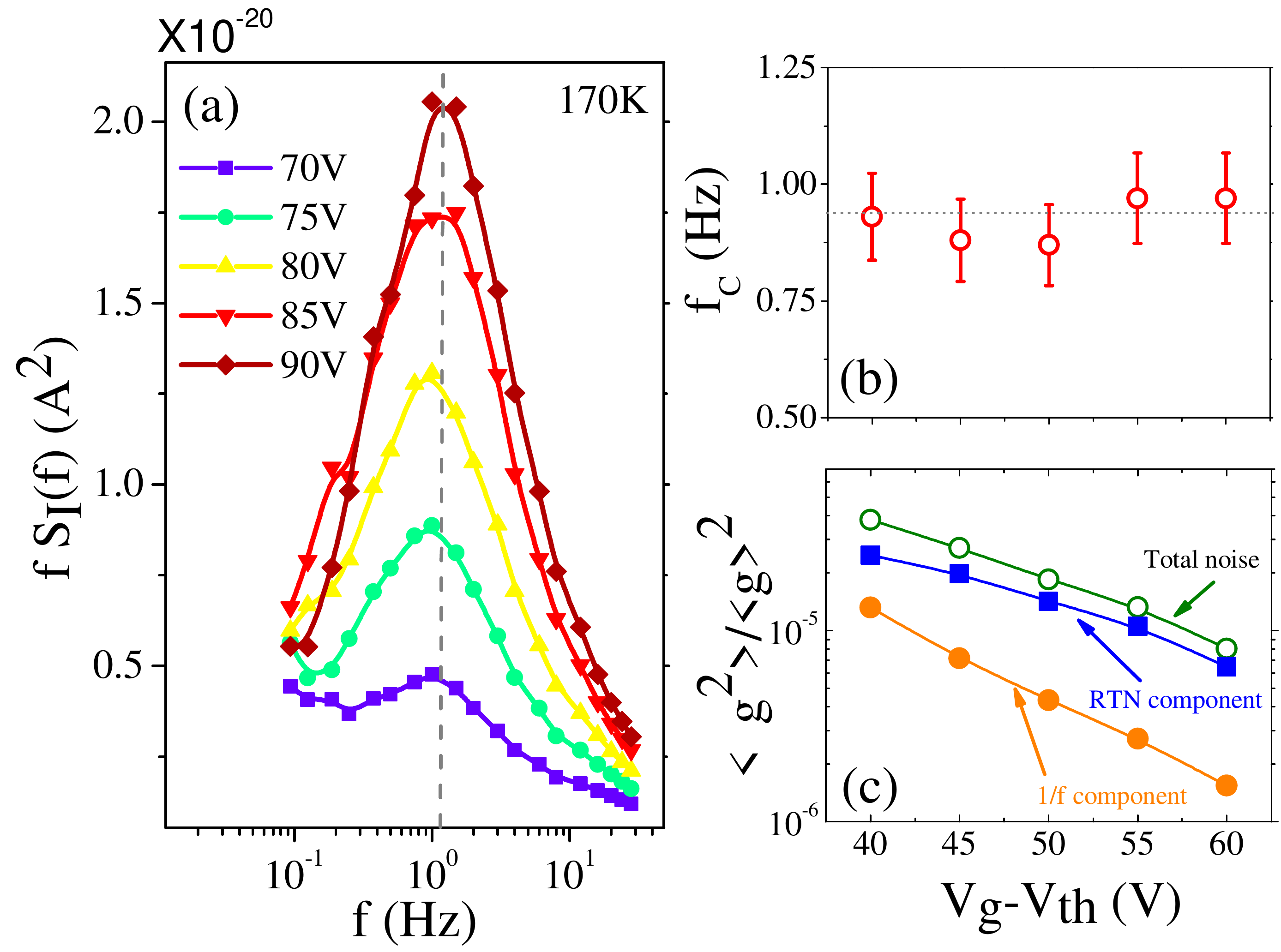}
		\small{{\caption{(a) Plot of $fS_I$(f) versus frequency $f$ at different gate-bias voltage $V_{g}$ across the sample D1. The vertical gray dashed line is a visual guide indicating the positions of the Lorentzian peaks in the spectra. (b) Plot of  $f_{c}$ as a function of $V_{g}$-$V_{th}$ extracted from the PSD in (a) using Eqn.~\ref{Eqn:2level}. The dotted line is a  guide to the eye. (c) Plots of the relative variance of conductance fluctuations $\mathcal{G}_{var}$, along with its RTN component and $1/f$-component versus $V_{g}$-$V_{th}$.  The solid lines are  guides to the eye. The data were all taken at  $T$ = 170~K and $V_{ds}$ = 5~mV.  \label{fig:8}}}}
	\end{center}
\end{figure}

In Fig.~\ref{fig:7}(a) we present the $V_{ds}$ dependence of the quantity $fS_I(f)$ measured at $T$ = 175~K and $V_g$ = 90~V for the device D1. We see that the  form of the PSD is independent of $V_{ds}$. To make this observation quantitative, we plot in Fig.~\ref{fig:7}(b) the dependence of $f_C$ on $V_{ds}$ extracted from these plots using Eqn.~\ref{Eqn:2level}. The fact that $f_c$ is independent of  $V_{ds}$ within experimental uncertainties shows that this time-scale is intrinsic to the sample~\cite{hooge1981experimental}. 

The PSD, $S_I(f)$  can be integrated over the frequency bandwidth of measurement  to obtain the relative variance of conductance fluctuations, $\mathcal{G}_{var}$ at a fixed $T$ and $V_g$: 
 \begin{eqnarray}
\mathcal{G}_{var} \equiv  \frac{\langle\delta g^2\rangle}{ \langle g\rangle^2}= \frac{\langle\delta I_{ds}^2\rangle}{ \langle I_{ds}\rangle^2} =\frac{1}{\langle I_{ds}\rangle^2}\int_{0.03125}^{28}{S_I(f)df}.
\end{eqnarray}
The relative variance of conductance fluctuations, $\mathcal{G}_{var}$ was  found to be independent of  $V_{ds}$ at all $T$ and $V_g$  confirming  that the noise arises from conductance fluctuations in the MoS$_2$ channel and not from the contacts [for representative data, see Fig.~\ref{fig:7}(c)].

We measured the noise as a function of gate-bias voltage, $V_g$ -- the results obtained at $T$ = 170~K for the device D1 are plotted in Fig.~\ref{fig:8}(a). We find $f_C$ to be independent of $V_{g}$ (Fig.~\ref{fig:8}(b)) with-in experimental uncertainties. In Fig.~\ref{fig:8}(c) we have plotted $\mathcal{G}_{var}$ as a function of $V_g$-$V_{th}$. The total noise has been separated into its $1/f$-component and the RTN-component. At low values of $V_g$-$V_{th}$, the $1/f$-component noise contribution is comparable to that of the RTN-component while at higher $V_g$-$V_{th}$, the RTN-component dominates the measured conductance fluctuations. This motivated us to perform our noise measurements at high $V_g$ (90~V) so that the RTN component of the noise is easily resolvable.

\begin{figure}[t]
	\begin{center}
		\includegraphics[width=0.5\textwidth]{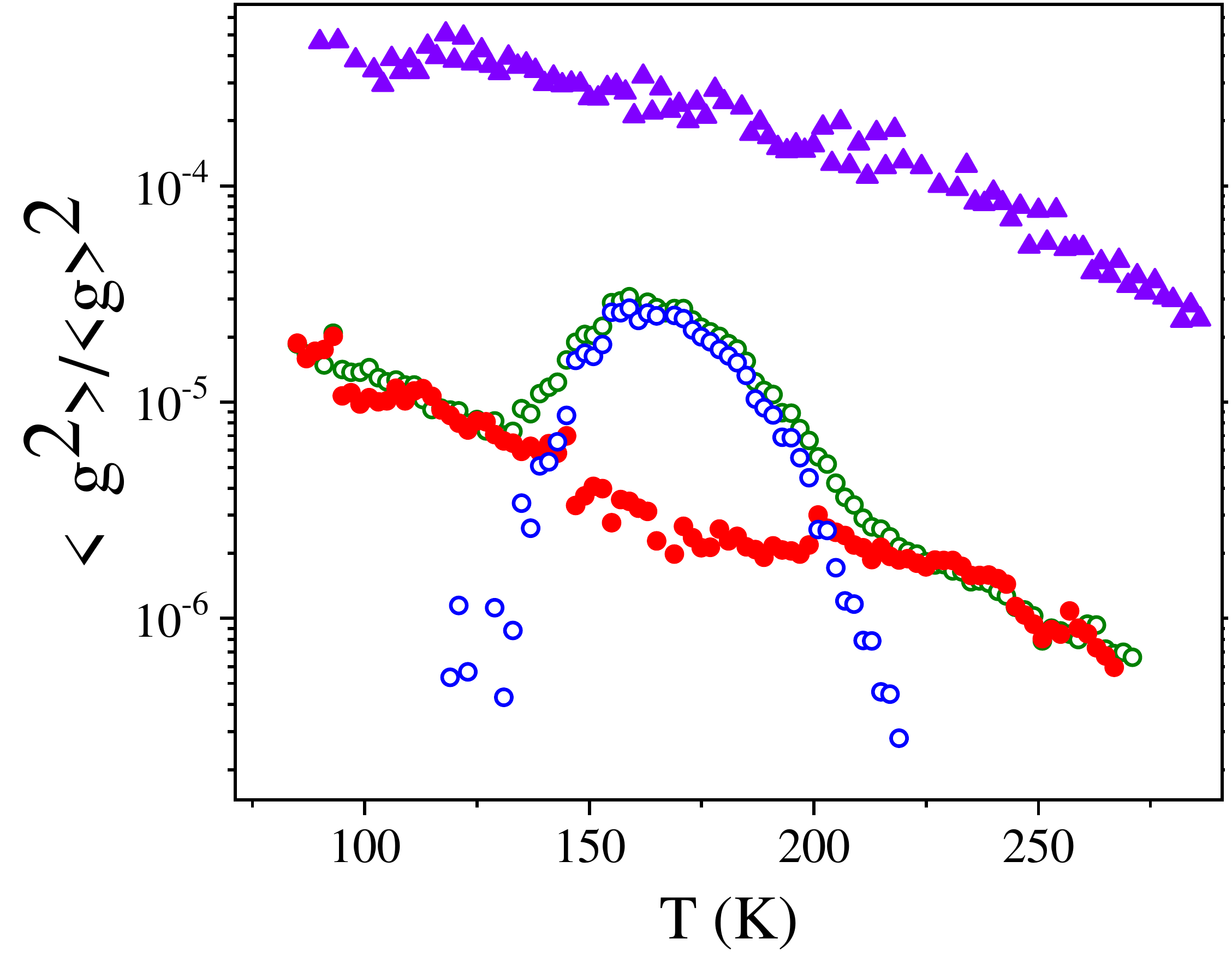}
		\small{{\caption{Comparison of the measured relative variance of conductance fluctuations $\mathcal{G}_{var}$ =${\langle\delta g^2\rangle}/{\langle g\rangle^2}$versus $T$  for device D1 (green-open circles) and D2 (purple-filled triangles). The red-filled circles and blue-open circles are respectively the contributions of the  $1/f$-component and the Lorentzian-component to  the measured noise  for the device D1. \label{fig:9}}}}
	\end{center}
\end{figure}

In Fig.~\ref{fig:9} we show a plot of  $\mathcal{G}_{var}$ versus $T$ for the two devices. The noise-data for device D1 (plotted in green open circles) has a prominent hump over the $T$ range ($\sim$140--190~K) coinciding with the regime where we observed RTN.  To appreciate this, we plot on the same graph the relative variance of conductance fluctuations arising from the $1/f$ component (red filled circles) as well as the Lorentzian component (blue open circles). It can be seen that the  increase in noise over the 140--190~K temperature range is entirely due to  two-level conductance fluctuations in the system. For comparison, we also add a plot of $\mathcal{G}_{var}$ versus $T$ for the unencapsulated device prepared on SiO$_2$-substrate, D2. The noise on SiO$_2$ substrate devices, is  more than two orders of magnitude larger than that of D1 and matches with previous reports of measured noise in MoS$_2$ by various groups ~\cite{renteria2014low,sangwan2013low}. Our work thus shows that encapsulation helps in significantly improving  the signal to noise ratio. 

\begin{figure}[t!]
	\begin{center}
		\includegraphics[width=0.5\textwidth]{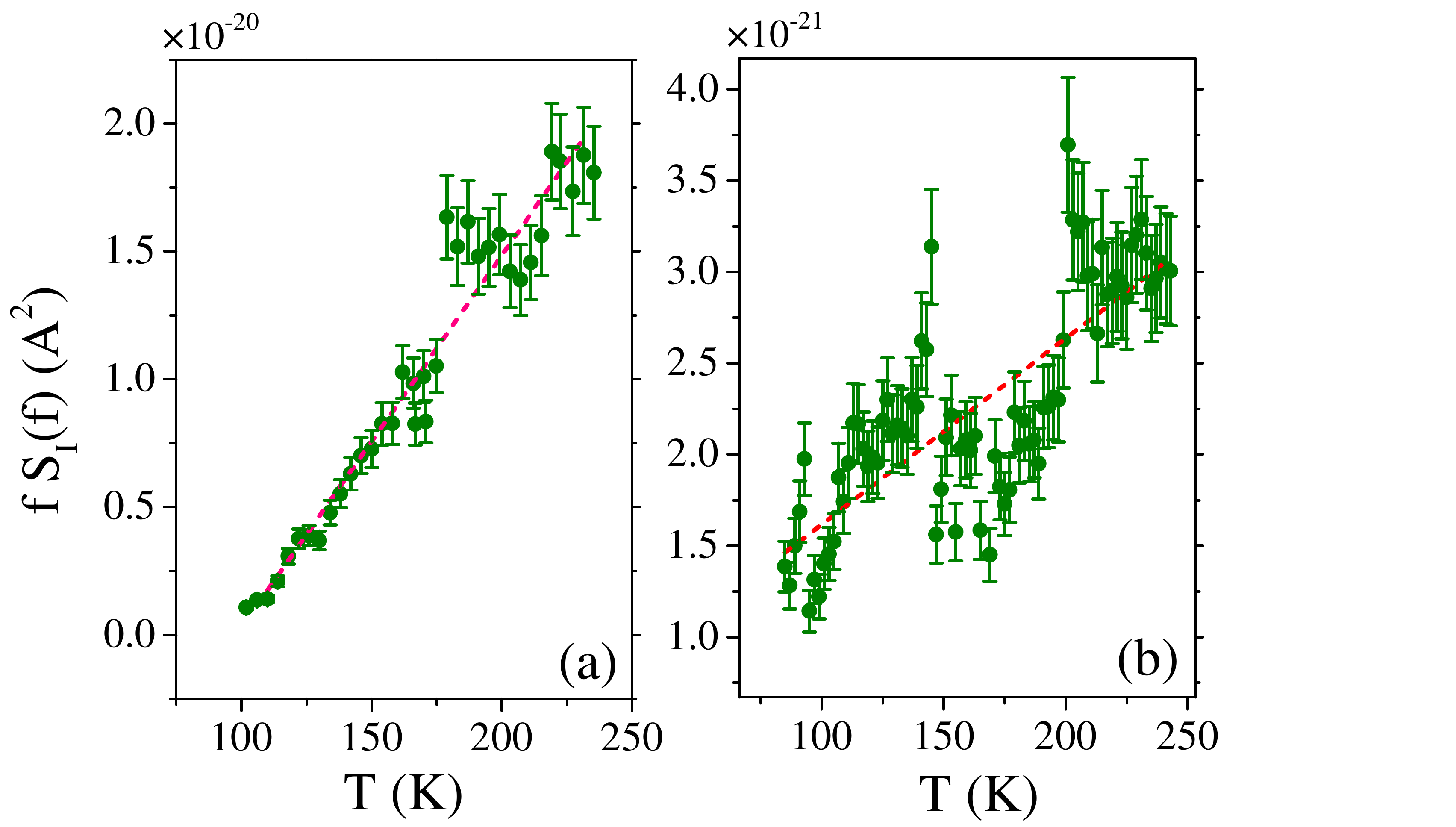}
		\small{{\caption{Plot of $fS_I(f)$  versus $T$ for the (a) on-SiO$_2$ substrate device D2, and (b) on-hBN HfO$_2$-encapsulated device D1. The red dashed line in both the plots are fits to Eqn.~\ref{Eqn:mcwhorter}.      \label{fig:10}}}}
	\end{center}
\end{figure}

A careful study of Fig.~\ref{fig:9} provides clues to the origin of the observed noise in this system. The temperature dependence of the relative variance of conductance fluctuations $\mathcal{G}_{var}$($T$) measured for the on-SiO$_2$ substrate device D2 closely resembles the $T$-dependence of the  $1/f$ component of $\mathcal{G}_{var}$ measured on D1. This indicates that these two noises have similar origins. The primary source of the $T$-dependence of noise in many semiconductor devices is   generation-recombination (GR)  noise due to  trapping-detrapping of charges at the gate dielectric-channel interfaces. This process can be quantified by the McWhorter model~\cite{mchorter,ghibaudo1991improved,na2014low}:
\begin{eqnarray}
N_{im} = fS_{I}(f)\langle R \rangle^2\frac{WLC^{2}}{e^{2}k_{B}}\frac{1}{T} 
\label{Eqn:mcwhorter}
\end{eqnarray} 
\noindent where $N_{im}$ is the areal-density of  trapped charges  per unit energy,   $W$ and $L$ are respectively the width and the length of the device-channel, $C$ is the gate-capacitance per unit area. $k_B$ is the Boltzmann constant and $e$ is the charge of the electron. Equation~\ref{Eqn:mcwhorter} predicts the linear dependence of $fS_I(f)$ on the temperature.  Fig.~\ref{fig:10} shows a plot of $fS_I(f)$ versus $T$ for both D1 and D2.  The plots are linear to within experimental uncertainties.  {From the slopes of these plots, the value of $N_{im}$ for device D2 was extracted to be $3.5\times 10^{12}$~cm$^{-2}$~eV$^{-1}$ which agrees with previously reported values for MoS$_2$ devices prepared on SiO$_2$ substrates}~\cite{doi:10.1021/acsnano.7b05520}. On the other hand, for the  {HfO$_2$ covered, on-hBN device} D1, $N_{im} = 1.8\times 10^{10}$~cm$^{-2}$~eV$^{-1}$, more than two orders of magnitude lower than that in the on-SiO$_2$ substrate device D2. 

The non-$1/f$ seen only in the encapsulated device has a different origin. The presence of RTN in the time-series of conductance fluctuations and the associated Lorentzian component  in the PSD indicates that  the noise originates from  random  charge fluctuations via transitions between two well-defined energy states separated by an energy barrier. We propose that in this case, these two levels correspond to the $\mathrm{S}$-vacancy  impurity band and the conduction band.  This is supported by the fact that the value of the activation energy, $E_a = 370$~meV extracted from the temperature dependence of the corner-frequency $f_C$  of the  Lorentzian component of the current fluctuations matches closely with the estimated position of the $\mathrm{S}$-vacancy  impurity band with respect to the conduction band edge~\cite{ghorbani2013defect}. Note that it was possible for us to detect this fluctuation-component only because of the two orders of noise  reduction made possible by the introduction of hBN between the MoS$_2$ and SiO$_2$  substrate.     

The HfO$_2$ layer has a two-fold effect on the noise. Firstly, being a high-k dielectric with a dielectric constant value of about 25, its presence screens the device from Coulomb scattering, and reduces the $1/f$ noise by orders of magnitude, enabling us to detect the RTN. Secondly, it acts as a capping layer that shields the MoS$_2$ from the ambient. We believe that this prevents the $\mathrm{S}$-vacancies from getting saturated by adsorbates, thus preserving the RTN. With the current data, we cannot distinguish between these two effects. Preliminary results obtained on devices fabricated on hBN without the HfO$_2$ capping layer had  higher on-off ratios, higher-mobilities and lower noise levels as compared to MoS$_2$ devices fabricated on SiO$_2$ without the HfO$_2$ capping   layer - however, we did not find any RTN in these devices. From these results, one can tentatively conclude that both the top- and bottom-layers are necessary to preserve the RTN.  This issue is currently under detailed investigation.

\begin{figure}[t]
	\begin{center}
		\includegraphics[width=0.5\textwidth]{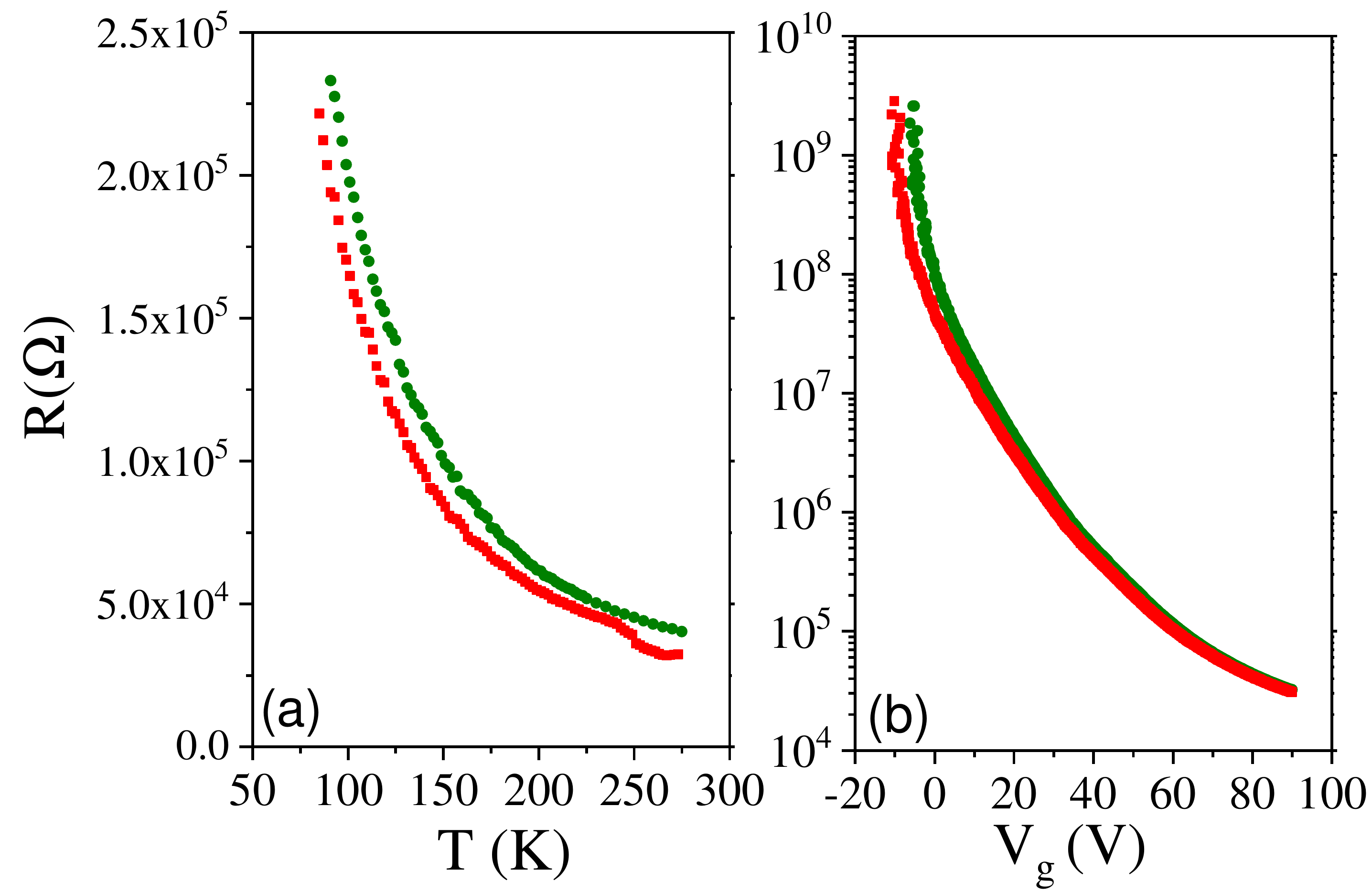}
		\small{{\caption {(a) Comparison between the  $R$ versus $T$ data at  $V_g =90$~V for  device D1 right after fabrication (red data points)  and several months as well as several thermal cycles  later (green data points). (b) Similar comparison of  $R$ versus $V_g$ data measured  at $T=270$~K for D1 in the pristine state and several months (and  thermal cycles) later.
					\label{fig:11}}}}
	\end{center}
\end{figure}

Finally, coming to the question of stability of the devices,  we have compared the $R$ versus $T$, $R$ versus $V_g$ and the noise measurements on  device D1 immediately after fabrication and after a gap of several months. The sample was thermally cycled several times during this period between 300~K and 77~K. As shown in Fig.~\ref{fig:11}, the temperature and $V_g$ dependence of the resistance of the device were quite reproducible. This is in sharp contrast to unencapsulated on-SiO$_2$ devices like D2  in which after a few days the channel and  contacts both degrade drastically making further  measurements impossible ~\cite{renteria2014low}. Similarly,  thermal cycling alters  the characteristics of such devices and makes the channel resistance unstable. The stability of the resistance of D1 over time period of months confirms that encapsulation between hBN and HfO$_2$ makes the device robust to thermal cycling and against degradation  with time.  

To conclude, in this paper we reported on detailed conductance fluctuation spectroscopy of high-quality MoS$_2$ devices  {encapsulated between hBN and HfO$_2$}. The presence of the high-$\kappa$ dielectric made the device extremely stable against environmental degradation enabling us to obtain reproducible data on the  same device for over 1 year. The hBN substrate helped bring down the conductance fluctuations by over two orders of magnitude as compared to similar devices on bare SiO$_2$ substrates. The low noise levels in our devices made it possible to detect the generation-recombination noise arising from charge fluctuation between the $\mathrm{S}$-vacancy levels in the MoS$_2$ band-gap and states at its conductance band edge. Our work establishes conduction fluctuation spectroscopy as a viable route to detect in-gap defect levels in low-dimensional semiconductors.

\begin{acknowledgments}
AB acknowledges funding from Nanomission and FIST program, Department of Science \& Technology (DST), government of India and the Indo-French Centre for the Promotion of Advanced Research (CEFIPRA) (Project No. 5304-F).
\end{acknowledgments}

\begin{thebibliography}{64}
	\expandafter\ifx\csname natexlab\endcsname\relax\def\natexlab#1{#1}\fi
	\expandafter\ifx\csname bibnamefont\endcsname\relax
	\def\bibnamefont#1{#1}\fi
	\expandafter\ifx\csname bibfnamefont\endcsname\relax
	\def\bibfnamefont#1{#1}\fi
	\expandafter\ifx\csname citenamefont\endcsname\relax
	\def\citenamefont#1{#1}\fi
	\expandafter\ifx\csname url\endcsname\relax
	\def\url#1{\texttt{#1}}\fi
	\expandafter\ifx\csname urlprefix\endcsname\relax\def\urlprefix{URL }\fi
	\providecommand{\bibinfo}[2]{#2}
	\providecommand{\eprint}[2][]{\url{#2}}
	
	\bibitem[{\citenamefont{Novoselov et~al.}(2004)\citenamefont{Novoselov, Geim,
			Morozov, Jiang, Zhang, Dubonos, Grigorieva, and
			Firsov}}]{novoselov2004electric}
	\bibinfo{author}{\bibfnamefont{K.~S.} \bibnamefont{Novoselov}},
	\bibinfo{author}{\bibfnamefont{A.~K.} \bibnamefont{Geim}},
	\bibinfo{author}{\bibfnamefont{S.~V.} \bibnamefont{Morozov}},
	\bibinfo{author}{\bibfnamefont{D.}~\bibnamefont{Jiang}},
	\bibinfo{author}{\bibfnamefont{Y.}~\bibnamefont{Zhang}},
	\bibinfo{author}{\bibfnamefont{S.~V.} \bibnamefont{Dubonos}},
	\bibinfo{author}{\bibfnamefont{I.~V.} \bibnamefont{Grigorieva}},
	\bibnamefont{and} \bibinfo{author}{\bibfnamefont{A.~A.}
		\bibnamefont{Firsov}}, \bibinfo{journal}{science}
	\textbf{\bibinfo{volume}{306}}, \bibinfo{pages}{666} (\bibinfo{year}{2004}).
	
	\bibitem[{\citenamefont{Mak et~al.}(2010)\citenamefont{Mak, Lee, Hone, Shan,
			and Heinz}}]{mak2010atomically}
	\bibinfo{author}{\bibfnamefont{K.~F.} \bibnamefont{Mak}},
	\bibinfo{author}{\bibfnamefont{C.}~\bibnamefont{Lee}},
	\bibinfo{author}{\bibfnamefont{J.}~\bibnamefont{Hone}},
	\bibinfo{author}{\bibfnamefont{J.}~\bibnamefont{Shan}}, \bibnamefont{and}
	\bibinfo{author}{\bibfnamefont{T.~F.} \bibnamefont{Heinz}},
	\bibinfo{journal}{Physical review letters} \textbf{\bibinfo{volume}{105}},
	\bibinfo{pages}{136805} (\bibinfo{year}{2010}).
	
	\bibitem[{\citenamefont{Wang et~al.}(2012)\citenamefont{Wang, Kalantar-Zadeh,
			Kis, Coleman, and Strano}}]{wang2012electronics}
	\bibinfo{author}{\bibfnamefont{Q.~H.} \bibnamefont{Wang}},
	\bibinfo{author}{\bibfnamefont{K.}~\bibnamefont{Kalantar-Zadeh}},
	\bibinfo{author}{\bibfnamefont{A.}~\bibnamefont{Kis}},
	\bibinfo{author}{\bibfnamefont{J.~N.} \bibnamefont{Coleman}},
	\bibnamefont{and} \bibinfo{author}{\bibfnamefont{M.~S.}
		\bibnamefont{Strano}}, \bibinfo{journal}{Nature nanotechnology}
	\textbf{\bibinfo{volume}{7}}, \bibinfo{pages}{699} (\bibinfo{year}{2012}).
	
	\bibitem[{\citenamefont{Radenovic et~al.}(2011)\citenamefont{Radenovic, Brivio,
			Giacometti, Kis et~al.}}]{radenovic2011single}
	\bibinfo{author}{\bibfnamefont{A.}~\bibnamefont{Radenovic}},
	\bibinfo{author}{\bibfnamefont{J.}~\bibnamefont{Brivio}},
	\bibinfo{author}{\bibfnamefont{V.}~\bibnamefont{Giacometti}},
	\bibinfo{author}{\bibfnamefont{A.}~\bibnamefont{Kis}}, \bibnamefont{et~al.},
	\bibinfo{journal}{Nat. Nanotechnol} \textbf{\bibinfo{volume}{6}},
	\bibinfo{pages}{147} (\bibinfo{year}{2011}).
	
	\bibitem[{\citenamefont{Eda and Maier}(2013)}]{eda2013two}
	\bibinfo{author}{\bibfnamefont{G.}~\bibnamefont{Eda}} \bibnamefont{and}
	\bibinfo{author}{\bibfnamefont{S.~A.} \bibnamefont{Maier}},
	\bibinfo{journal}{Acs Nano} \textbf{\bibinfo{volume}{7}},
	\bibinfo{pages}{5660} (\bibinfo{year}{2013}).
	
	\bibitem[{\citenamefont{Li et~al.}(2014)\citenamefont{Li, Wu, Yin, and
			Zhang}}]{doi:10.1021/ar4002312}
	\bibinfo{author}{\bibfnamefont{H.}~\bibnamefont{Li}},
	\bibinfo{author}{\bibfnamefont{J.}~\bibnamefont{Wu}},
	\bibinfo{author}{\bibfnamefont{Z.}~\bibnamefont{Yin}}, \bibnamefont{and}
	\bibinfo{author}{\bibfnamefont{H.}~\bibnamefont{Zhang}},
	\bibinfo{journal}{Accounts of Chemical Research}
	\textbf{\bibinfo{volume}{47}}, \bibinfo{pages}{1067} (\bibinfo{year}{2014}),
	\bibinfo{note}{pMID: 24697842}, \eprint{https://doi.org/10.1021/ar4002312},
	\urlprefix\url{https://doi.org/10.1021/ar4002312}.
	
	\bibitem[{\citenamefont{Hong et~al.}(2015)\citenamefont{Hong, Hu, Probert, Li,
			Lv, Yang, Gu, Mao, Feng, Xie et~al.}}]{hong2015exploring}
	\bibinfo{author}{\bibfnamefont{J.}~\bibnamefont{Hong}},
	\bibinfo{author}{\bibfnamefont{Z.}~\bibnamefont{Hu}},
	\bibinfo{author}{\bibfnamefont{M.}~\bibnamefont{Probert}},
	\bibinfo{author}{\bibfnamefont{K.}~\bibnamefont{Li}},
	\bibinfo{author}{\bibfnamefont{D.}~\bibnamefont{Lv}},
	\bibinfo{author}{\bibfnamefont{X.}~\bibnamefont{Yang}},
	\bibinfo{author}{\bibfnamefont{L.}~\bibnamefont{Gu}},
	\bibinfo{author}{\bibfnamefont{N.}~\bibnamefont{Mao}},
	\bibinfo{author}{\bibfnamefont{Q.}~\bibnamefont{Feng}},
	\bibinfo{author}{\bibfnamefont{L.}~\bibnamefont{Xie}}, \bibnamefont{et~al.},
	\bibinfo{journal}{Nature communications} \textbf{\bibinfo{volume}{6}},
	\bibinfo{pages}{6293} (\bibinfo{year}{2015}).
	
	\bibitem[{\citenamefont{Ong and Fischetti}(2013)}]{ong2013mobility}
	\bibinfo{author}{\bibfnamefont{Z.-Y.} \bibnamefont{Ong}} \bibnamefont{and}
	\bibinfo{author}{\bibfnamefont{M.~V.} \bibnamefont{Fischetti}},
	\bibinfo{journal}{Physical Review B} \textbf{\bibinfo{volume}{88}},
	\bibinfo{pages}{165316} (\bibinfo{year}{2013}).
	
	\bibitem[{\citenamefont{Qiu et~al.}(2013)\citenamefont{Qiu, Xu, Wang, Ren, Nan,
			Ni, Chen, Yuan, Miao, Song et~al.}}]{qiu2013hopping}
	\bibinfo{author}{\bibfnamefont{H.}~\bibnamefont{Qiu}},
	\bibinfo{author}{\bibfnamefont{T.}~\bibnamefont{Xu}},
	\bibinfo{author}{\bibfnamefont{Z.}~\bibnamefont{Wang}},
	\bibinfo{author}{\bibfnamefont{W.}~\bibnamefont{Ren}},
	\bibinfo{author}{\bibfnamefont{H.}~\bibnamefont{Nan}},
	\bibinfo{author}{\bibfnamefont{Z.}~\bibnamefont{Ni}},
	\bibinfo{author}{\bibfnamefont{Q.}~\bibnamefont{Chen}},
	\bibinfo{author}{\bibfnamefont{S.}~\bibnamefont{Yuan}},
	\bibinfo{author}{\bibfnamefont{F.}~\bibnamefont{Miao}},
	\bibinfo{author}{\bibfnamefont{F.}~\bibnamefont{Song}}, \bibnamefont{et~al.},
	\bibinfo{journal}{Nature communications} \textbf{\bibinfo{volume}{4}},
	\bibinfo{pages}{2642} (\bibinfo{year}{2013}).
	
	\bibitem[{\citenamefont{Zhu et~al.}(2014)\citenamefont{Zhu, Low, Lee, Wang,
			Farmer, Kong, Xia, and Avouris}}]{zhu2014electronic}
	\bibinfo{author}{\bibfnamefont{W.}~\bibnamefont{Zhu}},
	\bibinfo{author}{\bibfnamefont{T.}~\bibnamefont{Low}},
	\bibinfo{author}{\bibfnamefont{Y.-H.} \bibnamefont{Lee}},
	\bibinfo{author}{\bibfnamefont{H.}~\bibnamefont{Wang}},
	\bibinfo{author}{\bibfnamefont{D.~B.} \bibnamefont{Farmer}},
	\bibinfo{author}{\bibfnamefont{J.}~\bibnamefont{Kong}},
	\bibinfo{author}{\bibfnamefont{F.}~\bibnamefont{Xia}}, \bibnamefont{and}
	\bibinfo{author}{\bibfnamefont{P.}~\bibnamefont{Avouris}},
	\bibinfo{journal}{Nature Communications} \textbf{\bibinfo{volume}{5}},
	\bibinfo{pages}{3087} (\bibinfo{year}{2014}).
	
	\bibitem[{\citenamefont{McDonnell et~al.}(2014)\citenamefont{McDonnell, Addou,
			Buie, Wallace, and Hinkle}}]{mcdonnell2014defect}
	\bibinfo{author}{\bibfnamefont{S.}~\bibnamefont{McDonnell}},
	\bibinfo{author}{\bibfnamefont{R.}~\bibnamefont{Addou}},
	\bibinfo{author}{\bibfnamefont{C.}~\bibnamefont{Buie}},
	\bibinfo{author}{\bibfnamefont{R.~M.} \bibnamefont{Wallace}},
	\bibnamefont{and} \bibinfo{author}{\bibfnamefont{C.~L.}
		\bibnamefont{Hinkle}}, \bibinfo{journal}{ACS nano}
	\textbf{\bibinfo{volume}{8}}, \bibinfo{pages}{2880} (\bibinfo{year}{2014}).
	
	\bibitem[{\citenamefont{Tongay et~al.}(2013)\citenamefont{Tongay, Suh, Ataca,
			Fan, Luce, Kang, Liu, Ko, Raghunathanan, Zhou et~al.}}]{tongay2013defects}
	\bibinfo{author}{\bibfnamefont{S.}~\bibnamefont{Tongay}},
	\bibinfo{author}{\bibfnamefont{J.}~\bibnamefont{Suh}},
	\bibinfo{author}{\bibfnamefont{C.}~\bibnamefont{Ataca}},
	\bibinfo{author}{\bibfnamefont{W.}~\bibnamefont{Fan}},
	\bibinfo{author}{\bibfnamefont{A.}~\bibnamefont{Luce}},
	\bibinfo{author}{\bibfnamefont{J.~S.} \bibnamefont{Kang}},
	\bibinfo{author}{\bibfnamefont{J.}~\bibnamefont{Liu}},
	\bibinfo{author}{\bibfnamefont{C.}~\bibnamefont{Ko}},
	\bibinfo{author}{\bibfnamefont{R.}~\bibnamefont{Raghunathanan}},
	\bibinfo{author}{\bibfnamefont{J.}~\bibnamefont{Zhou}}, \bibnamefont{et~al.},
	\bibinfo{journal}{Scientific reports} \textbf{\bibinfo{volume}{3}},
	\bibinfo{pages}{2657} (\bibinfo{year}{2013}).
	
	\bibitem[{\citenamefont{Saigal and Ghosh}(2016)}]{saigal2016evidence}
	\bibinfo{author}{\bibfnamefont{N.}~\bibnamefont{Saigal}} \bibnamefont{and}
	\bibinfo{author}{\bibfnamefont{S.}~\bibnamefont{Ghosh}},
	\bibinfo{journal}{Applied Physics Letters} \textbf{\bibinfo{volume}{109}},
	\bibinfo{pages}{122105} (\bibinfo{year}{2016}).
	
	\bibitem[{\citenamefont{Lin et~al.}(2014)\citenamefont{Lin, Dumcenco, Komsa,
			Niimi, Krasheninnikov, Huang, and Suenaga}}]{lin2014properties}
	\bibinfo{author}{\bibfnamefont{Y.-C.} \bibnamefont{Lin}},
	\bibinfo{author}{\bibfnamefont{D.~O.} \bibnamefont{Dumcenco}},
	\bibinfo{author}{\bibfnamefont{H.-P.} \bibnamefont{Komsa}},
	\bibinfo{author}{\bibfnamefont{Y.}~\bibnamefont{Niimi}},
	\bibinfo{author}{\bibfnamefont{A.~V.} \bibnamefont{Krasheninnikov}},
	\bibinfo{author}{\bibfnamefont{Y.-S.} \bibnamefont{Huang}}, \bibnamefont{and}
	\bibinfo{author}{\bibfnamefont{K.}~\bibnamefont{Suenaga}},
	\bibinfo{journal}{Advanced materials} \textbf{\bibinfo{volume}{26}},
	\bibinfo{pages}{2857} (\bibinfo{year}{2014}).
	
	\bibitem[{\citenamefont{Zou et~al.}(2012)\citenamefont{Zou, Liu, and
			Yakobson}}]{zou2012predicting}
	\bibinfo{author}{\bibfnamefont{X.}~\bibnamefont{Zou}},
	\bibinfo{author}{\bibfnamefont{Y.}~\bibnamefont{Liu}}, \bibnamefont{and}
	\bibinfo{author}{\bibfnamefont{B.~I.} \bibnamefont{Yakobson}},
	\bibinfo{journal}{Nano letters} \textbf{\bibinfo{volume}{13}},
	\bibinfo{pages}{253} (\bibinfo{year}{2012}).
	
	\bibitem[{\citenamefont{Van Der~Zande et~al.}(2013)\citenamefont{Van Der~Zande,
			Huang, Chenet, Berkelbach, You, Lee, Heinz, Reichman, Muller, and
			Hone}}]{van2013grains}
	\bibinfo{author}{\bibfnamefont{A.~M.} \bibnamefont{Van Der~Zande}},
	\bibinfo{author}{\bibfnamefont{P.~Y.} \bibnamefont{Huang}},
	\bibinfo{author}{\bibfnamefont{D.~A.} \bibnamefont{Chenet}},
	\bibinfo{author}{\bibfnamefont{T.~C.} \bibnamefont{Berkelbach}},
	\bibinfo{author}{\bibfnamefont{Y.}~\bibnamefont{You}},
	\bibinfo{author}{\bibfnamefont{G.-H.} \bibnamefont{Lee}},
	\bibinfo{author}{\bibfnamefont{T.~F.} \bibnamefont{Heinz}},
	\bibinfo{author}{\bibfnamefont{D.~R.} \bibnamefont{Reichman}},
	\bibinfo{author}{\bibfnamefont{D.~A.} \bibnamefont{Muller}},
	\bibnamefont{and} \bibinfo{author}{\bibfnamefont{J.~C.} \bibnamefont{Hone}},
	\bibinfo{journal}{Nature materials} \textbf{\bibinfo{volume}{12}},
	\bibinfo{pages}{554} (\bibinfo{year}{2013}).
	
	\bibitem[{\citenamefont{Najmaei et~al.}(2013)\citenamefont{Najmaei, Liu, Zhou,
			Zou, Shi, Lei, Yakobson, Idrobo, Ajayan, and Lou}}]{najmaei2013vapour}
	\bibinfo{author}{\bibfnamefont{S.}~\bibnamefont{Najmaei}},
	\bibinfo{author}{\bibfnamefont{Z.}~\bibnamefont{Liu}},
	\bibinfo{author}{\bibfnamefont{W.}~\bibnamefont{Zhou}},
	\bibinfo{author}{\bibfnamefont{X.}~\bibnamefont{Zou}},
	\bibinfo{author}{\bibfnamefont{G.}~\bibnamefont{Shi}},
	\bibinfo{author}{\bibfnamefont{S.}~\bibnamefont{Lei}},
	\bibinfo{author}{\bibfnamefont{B.~I.} \bibnamefont{Yakobson}},
	\bibinfo{author}{\bibfnamefont{J.-C.} \bibnamefont{Idrobo}},
	\bibinfo{author}{\bibfnamefont{P.~M.} \bibnamefont{Ajayan}},
	\bibnamefont{and} \bibinfo{author}{\bibfnamefont{J.}~\bibnamefont{Lou}},
	\bibinfo{journal}{Nature materials} \textbf{\bibinfo{volume}{12}},
	\bibinfo{pages}{754} (\bibinfo{year}{2013}).
	
	\bibitem[{\citenamefont{Hosoki et~al.}(1992)\citenamefont{Hosoki, Hosaka, and
			Hasegawa}}]{hosoki1992surface}
	\bibinfo{author}{\bibfnamefont{S.}~\bibnamefont{Hosoki}},
	\bibinfo{author}{\bibfnamefont{S.}~\bibnamefont{Hosaka}}, \bibnamefont{and}
	\bibinfo{author}{\bibfnamefont{T.}~\bibnamefont{Hasegawa}},
	\bibinfo{journal}{Applied surface science} \textbf{\bibinfo{volume}{60}},
	\bibinfo{pages}{643} (\bibinfo{year}{1992}).
	
	\bibitem[{\citenamefont{Permana et~al.}(1992)\citenamefont{Permana, Lee, and
			Ng}}]{permana1992observation}
	\bibinfo{author}{\bibfnamefont{H.}~\bibnamefont{Permana}},
	\bibinfo{author}{\bibfnamefont{S.}~\bibnamefont{Lee}}, \bibnamefont{and}
	\bibinfo{author}{\bibfnamefont{K.~S.} \bibnamefont{Ng}},
	\bibinfo{journal}{Journal of Vacuum Science \& Technology B: Microelectronics
		and Nanometer Structures Processing, Measurement, and Phenomena}
	\textbf{\bibinfo{volume}{10}}, \bibinfo{pages}{2297} (\bibinfo{year}{1992}).
	
	\bibitem[{\citenamefont{Ha et~al.}(1994)\citenamefont{Ha, Roh, Park, Yi, and
			Lee}}]{ha1994scanning}
	\bibinfo{author}{\bibfnamefont{J.~S.} \bibnamefont{Ha}},
	\bibinfo{author}{\bibfnamefont{H.-S.} \bibnamefont{Roh}},
	\bibinfo{author}{\bibfnamefont{S.-J.} \bibnamefont{Park}},
	\bibinfo{author}{\bibfnamefont{J.-Y.} \bibnamefont{Yi}}, \bibnamefont{and}
	\bibinfo{author}{\bibfnamefont{E.-H.} \bibnamefont{Lee}},
	\bibinfo{journal}{Surface science} \textbf{\bibinfo{volume}{315}},
	\bibinfo{pages}{62} (\bibinfo{year}{1994}).
	
	\bibitem[{\citenamefont{Murata et~al.}(2001)\citenamefont{Murata, Kataoka, and
			Koma}}]{murata2001scanning}
	\bibinfo{author}{\bibfnamefont{H.}~\bibnamefont{Murata}},
	\bibinfo{author}{\bibfnamefont{K.}~\bibnamefont{Kataoka}}, \bibnamefont{and}
	\bibinfo{author}{\bibfnamefont{A.}~\bibnamefont{Koma}},
	\bibinfo{journal}{Surface science} \textbf{\bibinfo{volume}{478}},
	\bibinfo{pages}{131} (\bibinfo{year}{2001}).
	
	\bibitem[{\citenamefont{Kodama et~al.}(2010)\citenamefont{Kodama, Hasegawa,
			Okawa, Tsuruoka, Joachim, and Aono}}]{kodama2010electronic}
	\bibinfo{author}{\bibfnamefont{N.}~\bibnamefont{Kodama}},
	\bibinfo{author}{\bibfnamefont{T.}~\bibnamefont{Hasegawa}},
	\bibinfo{author}{\bibfnamefont{Y.}~\bibnamefont{Okawa}},
	\bibinfo{author}{\bibfnamefont{T.}~\bibnamefont{Tsuruoka}},
	\bibinfo{author}{\bibfnamefont{C.}~\bibnamefont{Joachim}}, \bibnamefont{and}
	\bibinfo{author}{\bibfnamefont{M.}~\bibnamefont{Aono}},
	\bibinfo{journal}{Japanese Journal of Applied Physics}
	\textbf{\bibinfo{volume}{49}}, \bibinfo{pages}{08LB01}
	(\bibinfo{year}{2010}).
	
	\bibitem[{\citenamefont{Addou et~al.}(2015)\citenamefont{Addou, Colombo, and
			Wallace}}]{addou2015surface}
	\bibinfo{author}{\bibfnamefont{R.}~\bibnamefont{Addou}},
	\bibinfo{author}{\bibfnamefont{L.}~\bibnamefont{Colombo}}, \bibnamefont{and}
	\bibinfo{author}{\bibfnamefont{R.~M.} \bibnamefont{Wallace}},
	\bibinfo{journal}{ACS applied materials \& interfaces}
	\textbf{\bibinfo{volume}{7}}, \bibinfo{pages}{11921} (\bibinfo{year}{2015}).
	
	\bibitem[{\citenamefont{Komsa et~al.}(2012)\citenamefont{Komsa, Kotakoski,
			Kurasch, Lehtinen, Kaiser, and Krasheninnikov}}]{komsa2012two}
	\bibinfo{author}{\bibfnamefont{H.-P.} \bibnamefont{Komsa}},
	\bibinfo{author}{\bibfnamefont{J.}~\bibnamefont{Kotakoski}},
	\bibinfo{author}{\bibfnamefont{S.}~\bibnamefont{Kurasch}},
	\bibinfo{author}{\bibfnamefont{O.}~\bibnamefont{Lehtinen}},
	\bibinfo{author}{\bibfnamefont{U.}~\bibnamefont{Kaiser}}, \bibnamefont{and}
	\bibinfo{author}{\bibfnamefont{A.~V.} \bibnamefont{Krasheninnikov}},
	\bibinfo{journal}{Physical review letters} \textbf{\bibinfo{volume}{109}},
	\bibinfo{pages}{035503} (\bibinfo{year}{2012}).
	
	\bibitem[{\citenamefont{Komsa et~al.}(2013)\citenamefont{Komsa, Kurasch,
			Lehtinen, Kaiser, and Krasheninnikov}}]{komsa2013point}
	\bibinfo{author}{\bibfnamefont{H.-P.} \bibnamefont{Komsa}},
	\bibinfo{author}{\bibfnamefont{S.}~\bibnamefont{Kurasch}},
	\bibinfo{author}{\bibfnamefont{O.}~\bibnamefont{Lehtinen}},
	\bibinfo{author}{\bibfnamefont{U.}~\bibnamefont{Kaiser}}, \bibnamefont{and}
	\bibinfo{author}{\bibfnamefont{A.~V.} \bibnamefont{Krasheninnikov}},
	\bibinfo{journal}{Physical Review B} \textbf{\bibinfo{volume}{88}},
	\bibinfo{pages}{035301} (\bibinfo{year}{2013}).
	
	\bibitem[{\citenamefont{Lu et~al.}(2014)\citenamefont{Lu, Li, Mao, Wang, and
			Andrei}}]{lu2014bandgap}
	\bibinfo{author}{\bibfnamefont{C.-P.} \bibnamefont{Lu}},
	\bibinfo{author}{\bibfnamefont{G.}~\bibnamefont{Li}},
	\bibinfo{author}{\bibfnamefont{J.}~\bibnamefont{Mao}},
	\bibinfo{author}{\bibfnamefont{L.-M.} \bibnamefont{Wang}}, \bibnamefont{and}
	\bibinfo{author}{\bibfnamefont{E.~Y.} \bibnamefont{Andrei}},
	\bibinfo{journal}{Nano letters} \textbf{\bibinfo{volume}{14}},
	\bibinfo{pages}{4628} (\bibinfo{year}{2014}).
	
	\bibitem[{\citenamefont{Huang et~al.}(2015)\citenamefont{Huang, Chen, Zhang,
			Quek, Chen, Li, Hsu, Chang, Zheng, Chen et~al.}}]{huang2015bandgap}
	\bibinfo{author}{\bibfnamefont{Y.~L.} \bibnamefont{Huang}},
	\bibinfo{author}{\bibfnamefont{Y.}~\bibnamefont{Chen}},
	\bibinfo{author}{\bibfnamefont{W.}~\bibnamefont{Zhang}},
	\bibinfo{author}{\bibfnamefont{S.~Y.} \bibnamefont{Quek}},
	\bibinfo{author}{\bibfnamefont{C.-H.} \bibnamefont{Chen}},
	\bibinfo{author}{\bibfnamefont{L.-J.} \bibnamefont{Li}},
	\bibinfo{author}{\bibfnamefont{W.-T.} \bibnamefont{Hsu}},
	\bibinfo{author}{\bibfnamefont{W.-H.} \bibnamefont{Chang}},
	\bibinfo{author}{\bibfnamefont{Y.~J.} \bibnamefont{Zheng}},
	\bibinfo{author}{\bibfnamefont{W.}~\bibnamefont{Chen}}, \bibnamefont{et~al.},
	\bibinfo{journal}{Nature communications} \textbf{\bibinfo{volume}{6}},
	\bibinfo{pages}{6298} (\bibinfo{year}{2015}).
	
	\bibitem[{\citenamefont{Vancs{\'o} et~al.}(2016)\citenamefont{Vancs{\'o},
			Magda, Pet{\H{o}}, Noh, Kim, Hwang, Bir{\'o}, and
			Tapaszt{\'o}}}]{vancso2016intrinsic}
	\bibinfo{author}{\bibfnamefont{P.}~\bibnamefont{Vancs{\'o}}},
	\bibinfo{author}{\bibfnamefont{G.~Z.} \bibnamefont{Magda}},
	\bibinfo{author}{\bibfnamefont{J.}~\bibnamefont{Pet{\H{o}}}},
	\bibinfo{author}{\bibfnamefont{J.-Y.} \bibnamefont{Noh}},
	\bibinfo{author}{\bibfnamefont{Y.-S.} \bibnamefont{Kim}},
	\bibinfo{author}{\bibfnamefont{C.}~\bibnamefont{Hwang}},
	\bibinfo{author}{\bibfnamefont{L.~P.} \bibnamefont{Bir{\'o}}},
	\bibnamefont{and}
	\bibinfo{author}{\bibfnamefont{L.}~\bibnamefont{Tapaszt{\'o}}},
	\bibinfo{journal}{Scientific reports} \textbf{\bibinfo{volume}{6}},
	\bibinfo{pages}{29726} (\bibinfo{year}{2016}).
	
	\bibitem[{\citenamefont{KC et~al.}(2014)\citenamefont{KC, Longo, Addou,
			Wallace, and Cho}}]{KC_2014}
	\bibinfo{author}{\bibfnamefont{S.}~\bibnamefont{KC}},
	\bibinfo{author}{\bibfnamefont{R.~C.} \bibnamefont{Longo}},
	\bibinfo{author}{\bibfnamefont{R.}~\bibnamefont{Addou}},
	\bibinfo{author}{\bibfnamefont{R.~M.} \bibnamefont{Wallace}},
	\bibnamefont{and} \bibinfo{author}{\bibfnamefont{K.}~\bibnamefont{Cho}},
	\bibinfo{journal}{Nanotechnology} \textbf{\bibinfo{volume}{25}},
	\bibinfo{pages}{375703} (\bibinfo{year}{2014}).
	
	\bibitem[{\citenamefont{Naik and Jain}(2018)}]{PhysRevMaterials.2.084002}
	\bibinfo{author}{\bibfnamefont{M.~H.} \bibnamefont{Naik}} \bibnamefont{and}
	\bibinfo{author}{\bibfnamefont{M.}~\bibnamefont{Jain}},
	\bibinfo{journal}{Phys. Rev. Materials} \textbf{\bibinfo{volume}{2}},
	\bibinfo{pages}{084002} (\bibinfo{year}{2018}).
	
	\bibitem[{\citenamefont{Krivosheeva et~al.}(2015)\citenamefont{Krivosheeva,
			Shaposhnikov, Borisenko, Lazzari, Waileong, Gusakova, and
			Tay}}]{krivosheeva2015theoretical}
	\bibinfo{author}{\bibfnamefont{A.~V.} \bibnamefont{Krivosheeva}},
	\bibinfo{author}{\bibfnamefont{V.~L.} \bibnamefont{Shaposhnikov}},
	\bibinfo{author}{\bibfnamefont{V.~E.} \bibnamefont{Borisenko}},
	\bibinfo{author}{\bibfnamefont{J.-L.} \bibnamefont{Lazzari}},
	\bibinfo{author}{\bibfnamefont{C.}~\bibnamefont{Waileong}},
	\bibinfo{author}{\bibfnamefont{J.}~\bibnamefont{Gusakova}}, \bibnamefont{and}
	\bibinfo{author}{\bibfnamefont{B.~K.} \bibnamefont{Tay}},
	\bibinfo{journal}{Journal of Semiconductors} \textbf{\bibinfo{volume}{36}},
	\bibinfo{pages}{122002} (\bibinfo{year}{2015}).
	
	\bibitem[{\citenamefont{Lin et~al.}(2016)\citenamefont{Lin, Carvalho, Kahn, Lv,
			Rao, Terrones, Pimenta, and Terrones}}]{lin2016defect}
	\bibinfo{author}{\bibfnamefont{Z.}~\bibnamefont{Lin}},
	\bibinfo{author}{\bibfnamefont{B.~R.} \bibnamefont{Carvalho}},
	\bibinfo{author}{\bibfnamefont{E.}~\bibnamefont{Kahn}},
	\bibinfo{author}{\bibfnamefont{R.}~\bibnamefont{Lv}},
	\bibinfo{author}{\bibfnamefont{R.}~\bibnamefont{Rao}},
	\bibinfo{author}{\bibfnamefont{H.}~\bibnamefont{Terrones}},
	\bibinfo{author}{\bibfnamefont{M.~A.} \bibnamefont{Pimenta}},
	\bibnamefont{and} \bibinfo{author}{\bibfnamefont{M.}~\bibnamefont{Terrones}},
	\bibinfo{journal}{2D Materials} \textbf{\bibinfo{volume}{3}},
	\bibinfo{pages}{022002} (\bibinfo{year}{2016}).
	
	\bibitem[{\citenamefont{Noh et~al.}(2014)\citenamefont{Noh, Kim, and
			Kim}}]{noh2014stability}
	\bibinfo{author}{\bibfnamefont{J.-Y.} \bibnamefont{Noh}},
	\bibinfo{author}{\bibfnamefont{H.}~\bibnamefont{Kim}}, \bibnamefont{and}
	\bibinfo{author}{\bibfnamefont{Y.-S.} \bibnamefont{Kim}},
	\bibinfo{journal}{Physical Review B} \textbf{\bibinfo{volume}{89}},
	\bibinfo{pages}{205417} (\bibinfo{year}{2014}).
	
	\bibitem[{\citenamefont{Dutta and Horn}(1981)}]{RevModPhys.53.497}
	\bibinfo{author}{\bibfnamefont{P.}~\bibnamefont{Dutta}} \bibnamefont{and}
	\bibinfo{author}{\bibfnamefont{P.~M.} \bibnamefont{Horn}},
	\bibinfo{journal}{Rev. Mod. Phys.} \textbf{\bibinfo{volume}{53}},
	\bibinfo{pages}{497} (\bibinfo{year}{1981}).
	
	\bibitem[{\citenamefont{Song et~al.}(2017)\citenamefont{Song, Joo, Neumann,
			Kim, and Lee}}]{song2017probing}
	\bibinfo{author}{\bibfnamefont{S.~H.} \bibnamefont{Song}},
	\bibinfo{author}{\bibfnamefont{M.-K.} \bibnamefont{Joo}},
	\bibinfo{author}{\bibfnamefont{M.}~\bibnamefont{Neumann}},
	\bibinfo{author}{\bibfnamefont{H.}~\bibnamefont{Kim}}, \bibnamefont{and}
	\bibinfo{author}{\bibfnamefont{Y.~H.} \bibnamefont{Lee}},
	\bibinfo{journal}{Nature communications} \textbf{\bibinfo{volume}{8}},
	\bibinfo{pages}{2121} (\bibinfo{year}{2017}).
	
	\bibitem[{\citenamefont{Kwon et~al.}(2014)\citenamefont{Kwon, Kang, Jang, Kim,
			and Grigoropoulos}}]{kwon2014analysis}
	\bibinfo{author}{\bibfnamefont{H.-J.} \bibnamefont{Kwon}},
	\bibinfo{author}{\bibfnamefont{H.}~\bibnamefont{Kang}},
	\bibinfo{author}{\bibfnamefont{J.}~\bibnamefont{Jang}},
	\bibinfo{author}{\bibfnamefont{S.}~\bibnamefont{Kim}}, \bibnamefont{and}
	\bibinfo{author}{\bibfnamefont{C.~P.} \bibnamefont{Grigoropoulos}},
	\bibinfo{journal}{Applied Physics Letters} \textbf{\bibinfo{volume}{104}},
	\bibinfo{pages}{083110} (\bibinfo{year}{2014}).
	
	\bibitem[{\citenamefont{Sangwan et~al.}(2013)\citenamefont{Sangwan, Arnold,
			Jariwala, Marks, Lauhon, and Hersam}}]{sangwan2013low}
	\bibinfo{author}{\bibfnamefont{V.~K.} \bibnamefont{Sangwan}},
	\bibinfo{author}{\bibfnamefont{H.~N.} \bibnamefont{Arnold}},
	\bibinfo{author}{\bibfnamefont{D.}~\bibnamefont{Jariwala}},
	\bibinfo{author}{\bibfnamefont{T.~J.} \bibnamefont{Marks}},
	\bibinfo{author}{\bibfnamefont{L.~J.} \bibnamefont{Lauhon}},
	\bibnamefont{and} \bibinfo{author}{\bibfnamefont{M.~C.}
		\bibnamefont{Hersam}}, \bibinfo{journal}{Nano letters}
	\textbf{\bibinfo{volume}{13}}, \bibinfo{pages}{4351} (\bibinfo{year}{2013}).
	
	\bibitem[{\citenamefont{Ghatak et~al.}(2014)\citenamefont{Ghatak, Mukherjee,
			Jain, Sarma, and Ghosh}}]{ghatak2014microscopic}
	\bibinfo{author}{\bibfnamefont{S.}~\bibnamefont{Ghatak}},
	\bibinfo{author}{\bibfnamefont{S.}~\bibnamefont{Mukherjee}},
	\bibinfo{author}{\bibfnamefont{M.}~\bibnamefont{Jain}},
	\bibinfo{author}{\bibfnamefont{D.}~\bibnamefont{Sarma}}, \bibnamefont{and}
	\bibinfo{author}{\bibfnamefont{A.}~\bibnamefont{Ghosh}},
	\bibinfo{journal}{APL Materials} \textbf{\bibinfo{volume}{2}},
	\bibinfo{pages}{092515} (\bibinfo{year}{2014}).
	
	\bibitem[{\citenamefont{Renteria et~al.}(2014)\citenamefont{Renteria, Samnakay,
			Rumyantsev, Jiang, Goli, Shur, and Balandin}}]{renteria2014low}
	\bibinfo{author}{\bibfnamefont{J.}~\bibnamefont{Renteria}},
	\bibinfo{author}{\bibfnamefont{R.}~\bibnamefont{Samnakay}},
	\bibinfo{author}{\bibfnamefont{S.}~\bibnamefont{Rumyantsev}},
	\bibinfo{author}{\bibfnamefont{C.}~\bibnamefont{Jiang}},
	\bibinfo{author}{\bibfnamefont{P.}~\bibnamefont{Goli}},
	\bibinfo{author}{\bibfnamefont{M.}~\bibnamefont{Shur}}, \bibnamefont{and}
	\bibinfo{author}{\bibfnamefont{A.}~\bibnamefont{Balandin}},
	\bibinfo{journal}{Applied Physics Letters} \textbf{\bibinfo{volume}{104}},
	\bibinfo{pages}{153104} (\bibinfo{year}{2014}).
	
	\bibitem[{\citenamefont{Na et~al.}(2014)\citenamefont{Na, Joo, Shin, Huh, Kim,
			Piao, Jin, Jang, Choi, Shim et~al.}}]{na2014low}
	\bibinfo{author}{\bibfnamefont{J.}~\bibnamefont{Na}},
	\bibinfo{author}{\bibfnamefont{M.-K.} \bibnamefont{Joo}},
	\bibinfo{author}{\bibfnamefont{M.}~\bibnamefont{Shin}},
	\bibinfo{author}{\bibfnamefont{J.}~\bibnamefont{Huh}},
	\bibinfo{author}{\bibfnamefont{J.-S.} \bibnamefont{Kim}},
	\bibinfo{author}{\bibfnamefont{M.}~\bibnamefont{Piao}},
	\bibinfo{author}{\bibfnamefont{J.-E.} \bibnamefont{Jin}},
	\bibinfo{author}{\bibfnamefont{H.-K.} \bibnamefont{Jang}},
	\bibinfo{author}{\bibfnamefont{H.~J.} \bibnamefont{Choi}},
	\bibinfo{author}{\bibfnamefont{J.~H.} \bibnamefont{Shim}},
	\bibnamefont{et~al.}, \bibinfo{journal}{Nanoscale}
	\textbf{\bibinfo{volume}{6}}, \bibinfo{pages}{433} (\bibinfo{year}{2014}).
	
	\bibitem[{\citenamefont{Qiu et~al.}(2012)\citenamefont{Qiu, Pan, Yao, Li, Shi,
			and Wang}}]{qiu2012electrical}
	\bibinfo{author}{\bibfnamefont{H.}~\bibnamefont{Qiu}},
	\bibinfo{author}{\bibfnamefont{L.}~\bibnamefont{Pan}},
	\bibinfo{author}{\bibfnamefont{Z.}~\bibnamefont{Yao}},
	\bibinfo{author}{\bibfnamefont{J.}~\bibnamefont{Li}},
	\bibinfo{author}{\bibfnamefont{Y.}~\bibnamefont{Shi}}, \bibnamefont{and}
	\bibinfo{author}{\bibfnamefont{X.}~\bibnamefont{Wang}},
	\bibinfo{journal}{Applied Physics Letters} \textbf{\bibinfo{volume}{100}},
	\bibinfo{pages}{123104} (\bibinfo{year}{2012}).
	
	\bibitem[{\citenamefont{Late et~al.}(2012)\citenamefont{Late, Liu, Matte,
			Dravid, and Rao}}]{doi:10.1021/nn301572c}
	\bibinfo{author}{\bibfnamefont{D.~J.} \bibnamefont{Late}},
	\bibinfo{author}{\bibfnamefont{B.}~\bibnamefont{Liu}},
	\bibinfo{author}{\bibfnamefont{H.~S. S.~R.} \bibnamefont{Matte}},
	\bibinfo{author}{\bibfnamefont{V.~P.} \bibnamefont{Dravid}},
	\bibnamefont{and} \bibinfo{author}{\bibfnamefont{C.~N.~R.}
		\bibnamefont{Rao}}, \bibinfo{journal}{ACS Nano} \textbf{\bibinfo{volume}{6}},
	\bibinfo{pages}{5635} (\bibinfo{year}{2012}), \bibinfo{note}{pMID: 22577885}.
	
	\bibitem[{\citenamefont{Kooyman and van Veen}(2008)}]{kooyman2008detrimental}
	\bibinfo{author}{\bibfnamefont{P.~J.} \bibnamefont{Kooyman}} \bibnamefont{and}
	\bibinfo{author}{\bibfnamefont{J.~R.} \bibnamefont{van Veen}},
	\bibinfo{journal}{Catalysis Today} \textbf{\bibinfo{volume}{130}},
	\bibinfo{pages}{135} (\bibinfo{year}{2008}).
	
	\bibitem[{\citenamefont{Dean et~al.}(2010)\citenamefont{Dean, Young, Meric,
			Lee, Wang, Sorgenfrei, Watanabe, Taniguchi, Kim, Shepard
			et~al.}}]{dean2010boron}
	\bibinfo{author}{\bibfnamefont{C.~R.} \bibnamefont{Dean}},
	\bibinfo{author}{\bibfnamefont{A.~F.} \bibnamefont{Young}},
	\bibinfo{author}{\bibfnamefont{I.}~\bibnamefont{Meric}},
	\bibinfo{author}{\bibfnamefont{C.}~\bibnamefont{Lee}},
	\bibinfo{author}{\bibfnamefont{L.}~\bibnamefont{Wang}},
	\bibinfo{author}{\bibfnamefont{S.}~\bibnamefont{Sorgenfrei}},
	\bibinfo{author}{\bibfnamefont{K.}~\bibnamefont{Watanabe}},
	\bibinfo{author}{\bibfnamefont{T.}~\bibnamefont{Taniguchi}},
	\bibinfo{author}{\bibfnamefont{P.}~\bibnamefont{Kim}},
	\bibinfo{author}{\bibfnamefont{K.~L.} \bibnamefont{Shepard}},
	\bibnamefont{et~al.}, \bibinfo{journal}{Nature nanotechnology}
	\textbf{\bibinfo{volume}{5}}, \bibinfo{pages}{722} (\bibinfo{year}{2010}).
	
	\bibitem[{\citenamefont{Lakshmi~Ganapathi
			et~al.}(2013)\citenamefont{Lakshmi~Ganapathi, Bhat, and
			Mohan}}]{lakshmi2013optimization}
	\bibinfo{author}{\bibfnamefont{K.}~\bibnamefont{Lakshmi~Ganapathi}},
	\bibinfo{author}{\bibfnamefont{N.}~\bibnamefont{Bhat}}, \bibnamefont{and}
	\bibinfo{author}{\bibfnamefont{S.}~\bibnamefont{Mohan}},
	\bibinfo{journal}{Applied Physics Letters} \textbf{\bibinfo{volume}{103}},
	\bibinfo{pages}{073105} (\bibinfo{year}{2013}).
	
	\bibitem[{\citenamefont{Ganapathi et~al.}(2014)\citenamefont{Ganapathi, Bhat,
			and Mohan}}]{ganapathi2014influence}
	\bibinfo{author}{\bibfnamefont{K.~L.} \bibnamefont{Ganapathi}},
	\bibinfo{author}{\bibfnamefont{N.}~\bibnamefont{Bhat}}, \bibnamefont{and}
	\bibinfo{author}{\bibfnamefont{S.}~\bibnamefont{Mohan}},
	\bibinfo{journal}{Semiconductor Science and Technology}
	\textbf{\bibinfo{volume}{29}}, \bibinfo{pages}{055007}
	(\bibinfo{year}{2014}).
	
	\bibitem[{\citenamefont{Kim et~al.}(2012)\citenamefont{Kim, Konar, Hwang, Lee,
			Lee, Yang, Jung, Kim, Yoo, Choi et~al.}}]{kim2012high}
	\bibinfo{author}{\bibfnamefont{S.}~\bibnamefont{Kim}},
	\bibinfo{author}{\bibfnamefont{A.}~\bibnamefont{Konar}},
	\bibinfo{author}{\bibfnamefont{W.-S.} \bibnamefont{Hwang}},
	\bibinfo{author}{\bibfnamefont{J.~H.} \bibnamefont{Lee}},
	\bibinfo{author}{\bibfnamefont{J.}~\bibnamefont{Lee}},
	\bibinfo{author}{\bibfnamefont{J.}~\bibnamefont{Yang}},
	\bibinfo{author}{\bibfnamefont{C.}~\bibnamefont{Jung}},
	\bibinfo{author}{\bibfnamefont{H.}~\bibnamefont{Kim}},
	\bibinfo{author}{\bibfnamefont{J.-B.} \bibnamefont{Yoo}},
	\bibinfo{author}{\bibfnamefont{J.-Y.} \bibnamefont{Choi}},
	\bibnamefont{et~al.}, \bibinfo{journal}{Nature communications}
	\textbf{\bibinfo{volume}{3}}, \bibinfo{pages}{1011} (\bibinfo{year}{2012}).
	
	\bibitem[{\citenamefont{Sangwan and Hersam}(2018)}]{sangwan2018electronic}
	\bibinfo{author}{\bibfnamefont{V.~K.} \bibnamefont{Sangwan}} \bibnamefont{and}
	\bibinfo{author}{\bibfnamefont{M.~C.} \bibnamefont{Hersam}},
	\bibinfo{journal}{Annual review of physical chemistry}
	\textbf{\bibinfo{volume}{69}}, \bibinfo{pages}{299} (\bibinfo{year}{2018}).
	
	\bibitem[{\citenamefont{Mao et~al.}(2017)\citenamefont{Mao, Shuai, Song, Wu,
			Dally, Zhou, Liu, Sun, Zhang, dela Cruz et~al.}}]{mao2017manipulation}
	\bibinfo{author}{\bibfnamefont{J.}~\bibnamefont{Mao}},
	\bibinfo{author}{\bibfnamefont{J.}~\bibnamefont{Shuai}},
	\bibinfo{author}{\bibfnamefont{S.}~\bibnamefont{Song}},
	\bibinfo{author}{\bibfnamefont{Y.}~\bibnamefont{Wu}},
	\bibinfo{author}{\bibfnamefont{R.}~\bibnamefont{Dally}},
	\bibinfo{author}{\bibfnamefont{J.}~\bibnamefont{Zhou}},
	\bibinfo{author}{\bibfnamefont{Z.}~\bibnamefont{Liu}},
	\bibinfo{author}{\bibfnamefont{J.}~\bibnamefont{Sun}},
	\bibinfo{author}{\bibfnamefont{Q.}~\bibnamefont{Zhang}},
	\bibinfo{author}{\bibfnamefont{C.}~\bibnamefont{dela Cruz}},
	\bibnamefont{et~al.}, \bibinfo{journal}{Proceedings of the National Academy
		of Sciences} \textbf{\bibinfo{volume}{114}}, \bibinfo{pages}{10548}
	(\bibinfo{year}{2017}).
	
	\bibitem[{\citenamefont{Liu et~al.}(2017)\citenamefont{Liu, Huang, Park, Li,
			Yang, Liu, Liang, Minari, and Noh}}]{liu2017unified}
	\bibinfo{author}{\bibfnamefont{C.}~\bibnamefont{Liu}},
	\bibinfo{author}{\bibfnamefont{K.}~\bibnamefont{Huang}},
	\bibinfo{author}{\bibfnamefont{W.-T.} \bibnamefont{Park}},
	\bibinfo{author}{\bibfnamefont{M.}~\bibnamefont{Li}},
	\bibinfo{author}{\bibfnamefont{T.}~\bibnamefont{Yang}},
	\bibinfo{author}{\bibfnamefont{X.}~\bibnamefont{Liu}},
	\bibinfo{author}{\bibfnamefont{L.}~\bibnamefont{Liang}},
	\bibinfo{author}{\bibfnamefont{T.}~\bibnamefont{Minari}}, \bibnamefont{and}
	\bibinfo{author}{\bibfnamefont{Y.-Y.} \bibnamefont{Noh}},
	\bibinfo{journal}{Materials Horizons} \textbf{\bibinfo{volume}{4}},
	\bibinfo{pages}{608} (\bibinfo{year}{2017}).
	
	\bibitem[{\citenamefont{Bid}(2006)}]{Aveekthesis}
	\bibinfo{author}{\bibfnamefont{A.}~\bibnamefont{Bid}}, Ph.D. thesis,
	\bibinfo{school}{Indian Institute of Science} (\bibinfo{year}{2006}).
	
	\bibitem[{\citenamefont{Daptary et~al.}(2018)\citenamefont{Daptary, Kumar,
			Dogra, and Bid}}]{daptary2018effect}
	\bibinfo{author}{\bibfnamefont{G.~N.} \bibnamefont{Daptary}},
	\bibinfo{author}{\bibfnamefont{P.}~\bibnamefont{Kumar}},
	\bibinfo{author}{\bibfnamefont{A.}~\bibnamefont{Dogra}}, \bibnamefont{and}
	\bibinfo{author}{\bibfnamefont{A.}~\bibnamefont{Bid}},
	\bibinfo{journal}{Physical Review B} \textbf{\bibinfo{volume}{98}},
	\bibinfo{pages}{035433} (\bibinfo{year}{2018}).
	
	\bibitem[{\citenamefont{Price}(1981)}]{price1981two}
	\bibinfo{author}{\bibfnamefont{P.}~\bibnamefont{Price}},
	\bibinfo{journal}{Annals of Physics} \textbf{\bibinfo{volume}{133}},
	\bibinfo{pages}{217} (\bibinfo{year}{1981}).
	
	\bibitem[{\citenamefont{Hooge}(1994)}]{hooge19941}
	\bibinfo{author}{\bibfnamefont{F.}~\bibnamefont{Hooge}}, \bibinfo{journal}{IEEE
		Transactions on Electron Devices} \textbf{\bibinfo{volume}{41}},
	\bibinfo{pages}{1926} (\bibinfo{year}{1994}).
	
	\bibitem[{\citenamefont{Ghosh et~al.}(2004)\citenamefont{Ghosh, Kar, Bid, and
			Raychaudhuri}}]{ghosh2004set}
	\bibinfo{author}{\bibfnamefont{A.}~\bibnamefont{Ghosh}},
	\bibinfo{author}{\bibfnamefont{S.}~\bibnamefont{Kar}},
	\bibinfo{author}{\bibfnamefont{A.}~\bibnamefont{Bid}}, \bibnamefont{and}
	\bibinfo{author}{\bibfnamefont{A.}~\bibnamefont{Raychaudhuri}},
	\bibinfo{journal}{arXiv preprint cond-mat/0402130}  (\bibinfo{year}{2004}).
	
	\bibitem[{\citenamefont{Scofield}(1987)}]{scofield1987ac}
	\bibinfo{author}{\bibfnamefont{J.~H.} \bibnamefont{Scofield}},
	\bibinfo{journal}{Review of scientific instruments}
	\textbf{\bibinfo{volume}{58}}, \bibinfo{pages}{985} (\bibinfo{year}{1987}).
	
	\bibitem[{\citenamefont{Du~Pr\'e}(1950)}]{PhysRev.78.615}
	\bibinfo{author}{\bibfnamefont{F.~K.} \bibnamefont{Du~Pr\'e}},
	\bibinfo{journal}{Phys. Rev.} \textbf{\bibinfo{volume}{78}},
	\bibinfo{pages}{615} (\bibinfo{year}{1950}).
	
	\bibitem[{\citenamefont{Kundu et~al.}(2017)\citenamefont{Kundu, Ray, Dolui,
			Bagwe, Choudhury, Krupanidhi, Das, Raychaudhuri, and Bid}}]{kundu2017quantum}
	\bibinfo{author}{\bibfnamefont{H.~K.} \bibnamefont{Kundu}},
	\bibinfo{author}{\bibfnamefont{S.}~\bibnamefont{Ray}},
	\bibinfo{author}{\bibfnamefont{K.}~\bibnamefont{Dolui}},
	\bibinfo{author}{\bibfnamefont{V.}~\bibnamefont{Bagwe}},
	\bibinfo{author}{\bibfnamefont{P.~R.} \bibnamefont{Choudhury}},
	\bibinfo{author}{\bibfnamefont{S.}~\bibnamefont{Krupanidhi}},
	\bibinfo{author}{\bibfnamefont{T.}~\bibnamefont{Das}},
	\bibinfo{author}{\bibfnamefont{P.}~\bibnamefont{Raychaudhuri}},
	\bibnamefont{and} \bibinfo{author}{\bibfnamefont{A.}~\bibnamefont{Bid}},
	\bibinfo{journal}{Physical review letters} \textbf{\bibinfo{volume}{119}},
	\bibinfo{pages}{226802} (\bibinfo{year}{2017}).
	
	\bibitem[{\citenamefont{Hung et~al.}(1990)\citenamefont{Hung, Ko, Hu, and
			Cheng}}]{hung1990random}
	\bibinfo{author}{\bibfnamefont{K.~K.} \bibnamefont{Hung}},
	\bibinfo{author}{\bibfnamefont{P.~K.} \bibnamefont{Ko}},
	\bibinfo{author}{\bibfnamefont{C.}~\bibnamefont{Hu}}, \bibnamefont{and}
	\bibinfo{author}{\bibfnamefont{Y.~C.} \bibnamefont{Cheng}},
	\bibinfo{journal}{IEEE electron device letters}
	\textbf{\bibinfo{volume}{11}}, \bibinfo{pages}{90} (\bibinfo{year}{1990}).
	
	\bibitem[{\citenamefont{Hooge et~al.}(1981)\citenamefont{Hooge, Kleinpenning,
			and Vandamme}}]{hooge1981experimental}
	\bibinfo{author}{\bibfnamefont{F.}~\bibnamefont{Hooge}},
	\bibinfo{author}{\bibfnamefont{T.}~\bibnamefont{Kleinpenning}},
	\bibnamefont{and} \bibinfo{author}{\bibfnamefont{L.}~\bibnamefont{Vandamme}},
	\bibinfo{journal}{Reports on progress in Physics}
	\textbf{\bibinfo{volume}{44}}, \bibinfo{pages}{479} (\bibinfo{year}{1981}).
	
	\bibitem[{\citenamefont{McWhorter}(1957)}]{mchorter}
	\bibinfo{author}{\bibfnamefont{A.~H.} \bibnamefont{McWhorter}},
	\emph{\bibinfo{title}{Semiconductor surface physics}}, \bibinfo{number}{207}
	(\bibinfo{publisher}{Univ. of Pennsylvania Press}, \bibinfo{year}{1957}).
	
	\bibitem[{\citenamefont{Ghibaudo et~al.}(1991)\citenamefont{Ghibaudo, Roux,
			Nguyen-Duc, Balestra, and Brini}}]{ghibaudo1991improved}
	\bibinfo{author}{\bibfnamefont{G.}~\bibnamefont{Ghibaudo}},
	\bibinfo{author}{\bibfnamefont{O.}~\bibnamefont{Roux}},
	\bibinfo{author}{\bibfnamefont{C.}~\bibnamefont{Nguyen-Duc}},
	\bibinfo{author}{\bibfnamefont{F.}~\bibnamefont{Balestra}}, \bibnamefont{and}
	\bibinfo{author}{\bibfnamefont{J.}~\bibnamefont{Brini}},
	\bibinfo{journal}{physica status solidi (a)} \textbf{\bibinfo{volume}{124}},
	\bibinfo{pages}{571} (\bibinfo{year}{1991}).
	
	\bibitem[{\citenamefont{Dubey et~al.}(2017)\citenamefont{Dubey, Lisi, Nayak,
			Herziger, Nguyen, Le~Quang, Cherkez, González, Dappe, Watanabe
			et~al.}}]{doi:10.1021/acsnano.7b05520}
	\bibinfo{author}{\bibfnamefont{S.}~\bibnamefont{Dubey}},
	\bibinfo{author}{\bibfnamefont{S.}~\bibnamefont{Lisi}},
	\bibinfo{author}{\bibfnamefont{G.}~\bibnamefont{Nayak}},
	\bibinfo{author}{\bibfnamefont{F.}~\bibnamefont{Herziger}},
	\bibinfo{author}{\bibfnamefont{V.-D.} \bibnamefont{Nguyen}},
	\bibinfo{author}{\bibfnamefont{T.}~\bibnamefont{Le~Quang}},
	\bibinfo{author}{\bibfnamefont{V.}~\bibnamefont{Cherkez}},
	\bibinfo{author}{\bibfnamefont{C.}~\bibnamefont{González}},
	\bibinfo{author}{\bibfnamefont{Y.~J.} \bibnamefont{Dappe}},
	\bibinfo{author}{\bibfnamefont{K.}~\bibnamefont{Watanabe}},
	\bibnamefont{et~al.}, \bibinfo{journal}{ACS Nano}
	\textbf{\bibinfo{volume}{11}}, \bibinfo{pages}{11206} (\bibinfo{year}{2017}),
	\bibinfo{note}{pMID: 28992415}.
	
	\bibitem[{\citenamefont{Ghorbani-Asl et~al.}(2013)\citenamefont{Ghorbani-Asl,
			Enyashin, Kuc, Seifert, and Heine}}]{ghorbani2013defect}
	\bibinfo{author}{\bibfnamefont{M.}~\bibnamefont{Ghorbani-Asl}},
	\bibinfo{author}{\bibfnamefont{A.~N.} \bibnamefont{Enyashin}},
	\bibinfo{author}{\bibfnamefont{A.}~\bibnamefont{Kuc}},
	\bibinfo{author}{\bibfnamefont{G.}~\bibnamefont{Seifert}}, \bibnamefont{and}
	\bibinfo{author}{\bibfnamefont{T.}~\bibnamefont{Heine}},
	\bibinfo{journal}{Physical Review B} \textbf{\bibinfo{volume}{88}},
	\bibinfo{pages}{245440} (\bibinfo{year}{2013}).
	
\end{thebibliography}

\end{document}